\documentclass[zpreprint,zbstpl]{zeus_paper}
\usepackage[english]{babel}
\newcommand{\be}{\begin{equation}} 
\newcommand{\ee}{\end{equation}}
\newcommand{\bea}{\begin{eqnarray}}
\newcommand{\eea}{\end{eqnarray}}

\newcommand{\ddif}[3]{\frac{d^{2}#1}{d#2 d#3}}
\newcommand{\Zdetdesc}{%
A detailed description of the ZEUS detector can be found 
elsewhere~\cite{zeus:1993:bluebook}. A brief outline of the 
components that are most relevant for this analysis is given
below.\xspace}

\newcommand{\Zsttdesc}[1]{%
The STT consisted of 48 sectors of two different sizes. Each sector
contained 192 (small sector) or 264 (large sector) straws of diameter
7.5 mm arranged into 3 layers. The sectors were trapezoidal in shape
and each subtended an azimuthal angle of $60^{\circ}$ -- 6 sectors
formed a so-called superlayer. A particle passing through the complete
detector traversed 8 superlayers, which were rotated around the beam
direction at angles of $30^{\circ}$ or $15^{\circ}$ to each other. The STT
covered the polar-angle region $5^{\circ}<\theta<23^{\circ}$.
}
\newcommand{\Zcaldesc}{%
The high-resolution uranium--scintillator calorimeter (CAL)~\citeCAL consisted 
of three parts: the forward (FCAL), the barrel (BCAL) and the rear (RCAL)
calorimeters. Each part was subdivided transversely into towers and
longitudinally into one electromagnetic section (EMC) and either one (in RCAL)
or two (in BCAL and FCAL) hadronic sections (HAC). The smallest subdivision of
the calorimeter is called a cell.  The CAL energy resolutions, as measured under
test-beam conditions, were $\sigma(E)/E=0.18/\sqrt{E}$ for electrons and
$\sigma(E)/E=0.35/\sqrt{E}$ for hadrons, with $E$ in $\Gev$.}

\newcommand{\Zlumidesc}[1]{%
The luminosity was measured using the Bethe-Heitler reaction
$ep\,\rightarrow\, e\gamma p$ by a luminosity detector which consisted
of independent lead--scintillator calorimeter~\cite{%
desy-92-066,*zfp:c63:391,*acpp:b32:2025%
}
and magnetic spectrometer~\cite{%
physics-0512153%
} systems. The fractional systematic
uncertainty on the measured luminosity was #1.}
\newcommand{\Zacknowledge}{%
We appreciate the contributions to the construction and maintenance of the 
ZEUS de- tector of many people who are not listed as authors. The HERA 
machine group and the DESY computing staff are especially acknowledged for 
their success in providing excel- lent operation of the collider and the 
data-analysis environment. We thank the DESY directorate for their strong 
support and encouragement.}

\chardef\usc=95
\chardef\til=126
\catcode`\@=11 
\DeclareRobustCommand\xdotspace{\futurelet\@let@token\@xdotspace}
\def\@xdotspace{%
  \ifx\@let@token.\else
  \ifx\@let@token\bgroup.\else
  \ifx\@let@token\egroup.\else
  \ifx\@let@token\/.\else
  \ifx\@let@token\ .\else
  \ifx\@let@token~.\else
  \ifx\@let@token!.\else
  \ifx\@let@token,.\else
  \ifx\@let@token:.\else
  \ifx\@let@token;.\else
  \ifx\@let@token?.\else
  \ifx\@let@token/.\else
  \ifx\@let@token'.\else
  \ifx\@let@token).\else
  \ifx\@let@token-.\else
  \ifx\@let@token\@xobeysp.\else
  \ifx\@let@token\space.\else
  \ifx\@let@token\@sptoken.\else
   .\space
   \fi\fi\fi\fi\fi\fi\fi\fi\fi\fi\fi\fi\fi\fi\fi\fi\fi\fi}
\catcode`\@=12 

\newcommand{\stru}[2]{%
   \relax\ifmmode\hbox{\vrule height#1 depth#2 width0pt}%
   \else\vrule height#1 depth#2 width0pt\fi}

\newcommand{\Ronum}[1]{\uppercase\expandafter{\romannumeral#1}}
\newcommand{\ronum}[1]{\expandafter{\romannumeral#1}}
\DeclareRobustCommand{\LaTeXZ}{%
  \LaTeX\kern-.05em4\kern-.1em
  {\raisebox{-0.2ex}{$\scriptstyle\text{ZEUS}$}}\xspace}

\newcommand{\fig}[1]{Fig.~\ref{fig-#1}}
\newcommand{\Fig}[1]{Figure~\ref{fig-#1}}

\DeclareMathAlphabet{\mathbf}{OT1}{cmr}{bx}{sl}
\newcommand{\eVdist}{\kern-0.06667em}

\newcommand{\Gev}{{\text{Ge}\eVdist\text{V\/}}}

\newcommand{\gev}{{\,\text{Ge}\eVdist\text{V\/}}}

\newcommand{\pb}{\,\text{pb}}

\newcommand{\pbi}{\,\text{pb}^{-1}}

\newcommand{\cm}{\,\text{cm}}

\newcommand{\slashfrac}[2]{%
  \raisebox{0.5ex}{\ensuremath #1}\kern-0.12em/\kern-0.08em
  \raisebox{-.8ex}{\ensuremath #2}}

\newcommand{\sqr}[3]{%
    {\vcenter{\hrule height.#3ex\hbox{\vrule width.#2ex height#1ex
     \kern#1ex\vrule width.#3ex}\hrule height.#2ex}}}

\catcode`\@=11 
\newcommand{\parenbar}{\mathpalette\p@renb@r}
\def\p@renb@r#1#2{\vbox{%
  \ifx#1\scriptscriptstyle \dimen@.7em\dimen@ii.2em\else
  \ifx#1\scriptstyle \dimen@.8em\dimen@ii.25em\else
  \dimen@1em\dimen@ii.4em\fi\fi \offinterlineskip
  \ialign{\hfill##\hfill\cr
    \vbox{\hrule width\dimen@ii}\cr
    \noalign{\vskip-.3ex}%
    \hbox to\dimen@{$\mathchar300\hfil\mathchar301$}\cr
    \noalign{\vskip-.3ex}%
    $#1#2$\cr}}}
\catcode`\@=12 

\newcommand{\sitil}{{\tilde\sigma}}

\newcommand{\DA}{{\rm DA}}

\newcommand{\IP}{{\rm I$\kern-0.01667em$P}\xspace}
\newcommand{\JB}{{\rm JB}}

\mathchardef\qsm=63
\mathchardef\pls=43
\mathchardef\mns=512
\mathchardef\plm=518
\mathchardef\eql=61
\mathchardef\smallleft=300
\mathchardef\smallright=301
\mathchardef\les=316
\mathchardef\gre=318
\mathchardef\leq=532
\mathchardef\grq=533

\catcode`\@=11 
\newcounter{pict@width}
\newcounter{pict@height}
\newlength{\pict@scale}
\newcommand{\psfigadd}[4]{%
\setcounter{pict@width}{1*\ratio{#2+\pict@scale/2}{\pict@scale}}
\setcounter{pict@height}{1*\ratio{#3+\pict@scale/2}{\pict@scale}}
\setlength{\unitlength}{\pict@scale}
\hbox to #2{\hspace{-\fill}\begin{picture}(\thepict@width,\thepict@height)
\put(0,0){\psfig{figure=#1,width=#2,height=#3,clip=}}
\SetScale{0.283466457}
\SetWidth{1.763889}
{#4}
\end{picture}}
}
\newcounter{pict@widthfst}
\newcounter{pict@widthscd}
\newcounter{pict@widthtot}
\newcommand{\psfigaddtwo}[7]{%
\setcounter{pict@widthfst}{1*\ratio{#2+\pict@scale/2}{\pict@scale}}
\setcounter{pict@widthscd}{1*\ratio{#2+#4+\pict@scale/2}{\pict@scale}}
\setcounter{pict@widthtot}{1*\ratio{#2+#4+#6+\pict@scale/2}{\pict@scale}}
\setcounter{pict@height}{1*\ratio{#3+\pict@scale/2}{\pict@scale}}
\setlength{\unitlength}{\pict@scale}
\hbox{\hspace{-\fill}\begin{picture}(\thepict@widthtot,\thepict@height)
\put(0,0){\psfig{figure=#1,width=#2,height=#3,clip=}}
\put(\thepict@widthscd,0){\psfig{figure=#5,width=#6,height=#3,clip=}}
\SetScale{0.283466457}
\SetWidth{1.763889}
{#7}
\end{picture}}
}
\newcommand{\psfigror}[4]{%
\setcounter{pict@width}{1*\ratio{#2+\pict@scale/2}{\pict@scale}}
\setcounter{pict@height}{1*\ratio{#3+\pict@scale/2}{\pict@scale}}
\setlength{\unitlength}{\pict@scale}
\hbox{\begin{picture}(\thepict@width,\thepict@height)
\put(0,\thepict@height){\psfig{figure=#1,width=#3,height=#2,clip=,angle=270}}
\SetScale{0.283466457}
\SetWidth{1.763889}
{#4}
\end{picture}}
}
\newcommand{\psfigrol}[4]{%
\setcounter{pict@width}{1*\ratio{#2+\pict@scale/2}{\pict@scale}}
\setcounter{pict@height}{1*\ratio{#3+\pict@scale/2}{\pict@scale}}
\setlength{\unitlength}{\pict@scale}
\hbox{\begin{picture}(\thepict@width,\thepict@height)
\put(0,0){\psfig{figure=#1,width=#3,height=#2,clip=,angle=90}}
\SetScale{0.283466457}
\SetWidth{1.763889}
{#4}
\end{picture}}
}
\catcode`\@=12 
\newlength\listtextwidth

\catcode`\@=11 
\newlength{\@tabfninsert}
\newlength{\@tabfnwidth}
\newcommand{\tabfootnote}[2]{%
  \setlength{\@tabfninsert}{0.8em}
  \setlength{\@tabfnwidth}{\textwidth}
  \addtolength{\@tabfnwidth}{-\@tabfninsert}
  \addtolength{\@tabfnwidth}{-0.4em}
  \noindent\makebox[\@tabfninsert][r]{\footnotesize$^{#1}$\hfil}\hfill%
  \parbox[t]{\@tabfnwidth}{\footnotesize #2\hfill}}
\catcode`\@=12 
\def\citeCTD{{\cite{%
nim:a279:290,*npps:b32:181,*nim:a338:254%
}}\xspace}
\def\citeMVD{{\cite{%
nim:a581:656%
}}\xspace}
\def\citeCAL{{\cite{%
nim:a309:77,*nim:a309:101,*nim:a321:356,*nim:a336:23%
}}\xspace}
\begin{document}
\prepnum{{}}
\title{
Measurement of high-$\bold{Q^{2}}$ neutral current\\
deep inelastic $\bold{e^+ p}$ scattering cross sections\\
with a longitudinally polarised positron beam at HERA \\
}                                                       
                    
\author{ZEUS Collaboration}
\date{}

\abstract{

Measurements of neutral current cross sections for deep inelastic
scattering in $e^{+}p$ collisions at HERA
with a longitudinally polarised positron beam are presented.
The single-differential cross-sections $d\sigma/dQ^{2}$,
$d\sigma/dx$ and $d\sigma/dy$
and the reduced 
cross-section $\tilde \sigma$
were measured in the kinematic region $Q^2 > 185 \gev^2$ and $y<0.9$,
where $Q^2$ is the four-momentum transfer squared, $x$ the Bjorken scaling variable, and $y$ the inelasticity
of the interaction.
The measurements were performed separately
for positively and negatively polarised positron beams.
The measurements are based on an integrated luminosity of $135.5\pbi$
collected with the ZEUS detector in 2006 and 2007
at a centre-of-mass energy of $318 \gev$.
The structure functions $\tilde{F_3}$ and $F_3^{\gamma Z}$
were determined by combining the $e^+ p$ results presented in this paper
with previously published $e^ - p$ neutral current results.
The asymmetry parameter $A^+$ is used to demonstrate the
parity violation predicted in electroweak interactions.
The measurements are well described by 
the predictions of the Standard Model.
}

\makezeustitle

\pagenumbering{Roman}

                                                   %
                                                   %
\begin{center}
{                      \Large  The ZEUS Collaboration              }
\end{center}

{\small


        {\raggedright
H.~Abramowicz$^{45, ah}$, 
I.~Abt$^{35}$, 
L.~Adamczyk$^{13}$, 
M.~Adamus$^{54}$, 
R.~Aggarwal$^{7, c}$, 
S.~Antonelli$^{4}$, 
P.~Antonioli$^{3}$, 
A.~Antonov$^{33}$, 
M.~Arneodo$^{50}$, 
O.~Arslan$^{5}$
% ...
        }

\newpage


\makebox[3em]{$^{1}$}
\begin{minipage}[t]{14cm}
{\it Argonne National Laboratory, Argonne, Illinois 60439-4815, USA}~$^{A}$

\end{minipage}\\
\makebox[3em]{$^{2}$}
\begin{minipage}[t]{14cm}
{\it Andrews University, Berrien Springs, Michigan 49104-0380, USA}

\end{minipage}\\
\makebox[3em]{$^{3}$}
\begin{minipage}[t]{14cm}
{\it INFN Bologna, Bologna, Italy}~$^{B}$

\end{minipage}\\
\makebox[3em]{$^{4}$}
\begin{minipage}[t]{14cm}
{\it University and INFN Bologna, Bologna, Italy}~$^{B}$

\end{minipage}\\
\makebox[3em]{$^{5}$}
\begin{minipage}[t]{14cm}
{\it Physikalisches Institut der Universit\"at Bonn,
Bonn, Germany}~$^{C}$

\end{minipage}\\
\makebox[3em]{$^{6}$}
\begin{minipage}[t]{14cm}
{\it H.H.~Wills Physics Laboratory, University of Bristol,
Bristol, United Kingdom}~$^{D}$

\end{minipage}\\
\makebox[3em]{$^{7}$}
\begin{minipage}[t]{14cm}
{\it Panjab University, Department of Physics, Chandigarh, India}

\end{minipage}\\
\makebox[3em]{$^{8}$}
\begin{minipage}[t]{14cm}
{\it Calabria University,
Physics Department and INFN, Cosenza, Italy}~$^{B}$

\end{minipage}\\
\makebox[3em]{$^{9}$}
\begin{minipage}[t]{14cm}
{\it Institute for Universe and Elementary Particles, Chonnam National University,\\
Kwangju, South Korea}

\end{minipage}\\
\makebox[3em]{$^{10}$}
\begin{minipage}[t]{14cm}
{\it Jabatan Fizik, Universiti Malaya, 50603 Kuala Lumpur, Malaysia}~$^{E}$

\end{minipage}\\
\makebox[3em]{$^{11}$}
\begin{minipage}[t]{14cm}
{\it Nevis Laboratories, Columbia University, Irvington on Hudson,
New York 10027, USA}~$^{F}$

\end{minipage}\\
\makebox[3em]{$^{12}$}
\begin{minipage}[t]{14cm}
{\it The Henryk Niewodniczanski Institute of Nuclear Physics, Polish Academy of \\
Sciences, Krakow, Poland}~$^{G}$

\end{minipage}\\
\makebox[3em]{$^{13}$}
\begin{minipage}[t]{14cm}
{\it AGH-University of Science and Technology, Faculty of Physics and Applied Computer
Science, Krakow, Poland}~$^{H}$

\end{minipage}\\
\makebox[3em]{$^{14}$}
\begin{minipage}[t]{14cm}
{\it Department of Physics, Jagellonian University, Cracow, Poland}

\end{minipage}\\
\makebox[3em]{$^{15}$}
\begin{minipage}[t]{14cm}
{\it Deutsches Elektronen-Synchrotron DESY, Hamburg, Germany}

\end{minipage}\\
\makebox[3em]{$^{16}$}
\begin{minipage}[t]{14cm}
{\it Deutsches Elektronen-Synchrotron DESY, Zeuthen, Germany}

\end{minipage}\\
\makebox[3em]{$^{17}$}
\begin{minipage}[t]{14cm}
{\it INFN Florence, Florence, Italy}~$^{B}$

\end{minipage}\\
\makebox[3em]{$^{18}$}
\begin{minipage}[t]{14cm}
{\it University and INFN Florence, Florence, Italy}~$^{B}$

\end{minipage}\\
\makebox[3em]{$^{19}$}
\begin{minipage}[t]{14cm}
{\it Fakult\"at f\"ur Physik der Universit\"at Freiburg i.Br.,
Freiburg i.Br., Germany}

\end{minipage}\\
\makebox[3em]{$^{20}$}
\begin{minipage}[t]{14cm}
{\it School of Physics and Astronomy, University of Glasgow,
Glasgow, United Kingdom}~$^{D}$

\end{minipage}\\
\makebox[3em]{$^{21}$}
\begin{minipage}[t]{14cm}
{\it Department of Engineering in Management and Finance, Univ. of
the Aegean, Chios, Greece}

\end{minipage}\\
\makebox[3em]{$^{22}$}
\begin{minipage}[t]{14cm}
{\it Hamburg University, Institute of Experimental Physics, Hamburg,
Germany}~$^{I}$

\end{minipage}\\
\makebox[3em]{$^{23}$}
\begin{minipage}[t]{14cm}
{\it Imperial College London, High Energy Nuclear Physics Group,
London, United Kingdom}~$^{D}$

\end{minipage}\\
\makebox[3em]{$^{24}$}
\begin{minipage}[t]{14cm}
{\it Institute of Particle and Nuclear Studies, KEK,
Tsukuba, Japan}~$^{J}$

\end{minipage}\\
\makebox[3em]{$^{25}$}
\begin{minipage}[t]{14cm}
{\it Institute of Physics and Technology of Ministry of Education and
Science of Kazakhstan, Almaty, Kazakhstan}

\end{minipage}\\
\makebox[3em]{$^{26}$}
\begin{minipage}[t]{14cm}
{\it Institute for Nuclear Research, National Academy of Sciences, Kyiv, Ukraine}

\end{minipage}\\
\makebox[3em]{$^{27}$}
\begin{minipage}[t]{14cm}
{\it Department of Nuclear Physics, National Taras Shevchenko University of Kyiv, Kyiv, Ukraine}

\end{minipage}\\
\makebox[3em]{$^{28}$}
\begin{minipage}[t]{14cm}
{\it Kyungpook National University, Center for High Energy Physics, Daegu,
South Korea}~$^{K}$

\end{minipage}\\
\makebox[3em]{$^{29}$}
\begin{minipage}[t]{14cm}
{\it Institut de Physique Nucl\'{e}aire, Universit\'{e} Catholique de Louvain, Louvain-la-Neuve,\\
Belgium}~$^{L}$

\end{minipage}\\
\makebox[3em]{$^{30}$}
\begin{minipage}[t]{14cm}
{\it Departamento de F\'{\i}sica Te\'orica, Universidad Aut\'onoma
de Madrid, Madrid, Spain}~$^{M}$

\end{minipage}\\
\makebox[3em]{$^{31}$}
\begin{minipage}[t]{14cm}
{\it Department of Physics, McGill University,
Montr\'eal, Qu\'ebec, Canada H3A 2T8}~$^{N}$

\end{minipage}\\
\makebox[3em]{$^{32}$}
\begin{minipage}[t]{14cm}
{\it Meiji Gakuin University, Faculty of General Education,
Yokohama, Japan}~$^{J}$

\end{minipage}\\
\makebox[3em]{$^{33}$}
\begin{minipage}[t]{14cm}
{\it Moscow Engineering Physics Institute, Moscow, Russia}~$^{O}$

\end{minipage}\\
\makebox[3em]{$^{34}$}
\begin{minipage}[t]{14cm}
{\it Lomonosov Moscow State University, Skobeltsyn Institute of Nuclear Physics,
Moscow, Russia}~$^{P}$

\end{minipage}\\
\makebox[3em]{$^{35}$}
\begin{minipage}[t]{14cm}
{\it Max-Planck-Institut f\"ur Physik, M\"unchen, Germany}

\end{minipage}\\
\makebox[3em]{$^{36}$}
\begin{minipage}[t]{14cm}
{\it NIKHEF and University of Amsterdam, Amsterdam, Netherlands}~$^{Q}$

\end{minipage}\\
\makebox[3em]{$^{37}$}
\begin{minipage}[t]{14cm}
{\it Physics Department, Ohio State University,
Columbus, Ohio 43210, USA}~$^{A}$

\end{minipage}\\
\makebox[3em]{$^{38}$}
\begin{minipage}[t]{14cm}
{\it Department of Physics, University of Oxford,
Oxford, United Kingdom}~$^{D}$

\end{minipage}\\
\makebox[3em]{$^{39}$}
\begin{minipage}[t]{14cm}
{\it INFN Padova, Padova, Italy}~$^{B}$

\end{minipage}\\
\makebox[3em]{$^{40}$}
\begin{minipage}[t]{14cm}
{\it Dipartimento di Fisica dell' Universit\`a and INFN,
Padova, Italy}~$^{B}$

\end{minipage}\\
\makebox[3em]{$^{41}$}
\begin{minipage}[t]{14cm}
{\it Department of Physics, Pennsylvania State University, University Park,\\
Pennsylvania 16802, USA}~$^{F}$

\end{minipage}\\
\makebox[3em]{$^{42}$}
\begin{minipage}[t]{14cm}
{\it Polytechnic University, Tokyo, Japan}~$^{J}$

\end{minipage}\\
\makebox[3em]{$^{43}$}
\begin{minipage}[t]{14cm}
{\it Dipartimento di Fisica, Universit\`a 'La Sapienza' and INFN,
Rome, Italy}~$^{B}$

\end{minipage}\\
\makebox[3em]{$^{44}$}
\begin{minipage}[t]{14cm}
{\it Rutherford Appleton Laboratory, Chilton, Didcot, Oxon,
United Kingdom}~$^{D}$

\end{minipage}\\
\makebox[3em]{$^{45}$}
\begin{minipage}[t]{14cm}
{\it Raymond and Beverly Sackler Faculty of Exact Sciences, School of Physics, \\
Tel Aviv University, Tel Aviv, Israel}~$^{R}$

\end{minipage}\\
\makebox[3em]{$^{46}$}
\begin{minipage}[t]{14cm}
{\it Department of Physics, Tokyo Institute of Technology,
Tokyo, Japan}~$^{J}$

\end{minipage}\\
\makebox[3em]{$^{47}$}
\begin{minipage}[t]{14cm}
{\it Department of Physics, University of Tokyo,
Tokyo, Japan}~$^{J}$

\end{minipage}\\
\makebox[3em]{$^{48}$}
\begin{minipage}[t]{14cm}
{\it Tokyo Metropolitan University, Department of Physics,
Tokyo, Japan}~$^{J}$

\end{minipage}\\
\makebox[3em]{$^{49}$}
\begin{minipage}[t]{14cm}
{\it Universit\`a di Torino and INFN, Torino, Italy}~$^{B}$

\end{minipage}\\
\makebox[3em]{$^{50}$}
\begin{minipage}[t]{14cm}
{\it Universit\`a del Piemonte Orientale, Novara, and INFN, Torino,
Italy}~$^{B}$

\end{minipage}\\
\makebox[3em]{$^{51}$}
\begin{minipage}[t]{14cm}
{\it Department of Physics, University of Toronto, Toronto, Ontario,
Canada M5S 1A7}~$^{N}$

\end{minipage}\\
\makebox[3em]{$^{52}$}
\begin{minipage}[t]{14cm}
{\it Physics and Astronomy Department, University College London,
London, United Kingdom}~$^{D}$

\end{minipage}\\
\makebox[3em]{$^{53}$}
\begin{minipage}[t]{14cm}
{\it Faculty of Physics, University of Warsaw, Warsaw, Poland}

\end{minipage}\\
\makebox[3em]{$^{54}$}
\begin{minipage}[t]{14cm}
{\it National Centre for Nuclear Research, Warsaw, Poland}

\end{minipage}\\
\makebox[3em]{$^{55}$}
\begin{minipage}[t]{14cm}
{\it Department of Particle Physics and Astrophysics, Weizmann
Institute, Rehovot, Israel}

\end{minipage}\\
\makebox[3em]{$^{56}$}
\begin{minipage}[t]{14cm}
{\it Department of Physics, University of Wisconsin, Madison,
Wisconsin 53706, USA}~$^{A}$

\end{minipage}\\
\makebox[3em]{$^{57}$}
\begin{minipage}[t]{14cm}
{\it Department of Physics, York University, Ontario, Canada M3J 1P3}~$^{N}$

\end{minipage}\\
\vspace{30em} \pagebreak[4]


\makebox[3ex]{$^{ A}$}
\begin{minipage}[t]{14cm}
 supported by the US Department of Energy\
\end{minipage}\\
\makebox[3ex]{$^{ B}$}
\begin{minipage}[t]{14cm}
 supported by the Italian National Institute for Nuclear Physics (INFN) \
\end{minipage}\\
\makebox[3ex]{$^{ C}$}
\begin{minipage}[t]{14cm}
 supported by the German Federal Ministry for Education and Research (BMBF), under
 contract No. 05 H09PDF\
\end{minipage}\\
\makebox[3ex]{$^{ D}$}
\begin{minipage}[t]{14cm}
 supported by the Science and Technology Facilities Council, UK\
\end{minipage}\\
\makebox[3ex]{$^{ E}$}
\begin{minipage}[t]{14cm}
 supported by an FRGS grant from the Malaysian government\
\end{minipage}\\
\makebox[3ex]{$^{ F}$}
\begin{minipage}[t]{14cm}
 supported by the US National Science Foundation. Any opinion,
 findings and conclusions or recommendations expressed in this material
 are those of the authors and do not necessarily reflect the views of the
 National Science Foundation.\
\end{minipage}\\
\makebox[3ex]{$^{ G}$}
\begin{minipage}[t]{14cm}
 supported by the Polish Ministry of Science and Higher Education as a scientific project No.
 DPN/N188/DESY/2009\
\end{minipage}\\
\makebox[3ex]{$^{ H}$}
\begin{minipage}[t]{14cm}
 supported by the Polish Ministry of Science and Higher Education and its grants
 for Scientific Research\
\end{minipage}\\
\makebox[3ex]{$^{ I}$}
\begin{minipage}[t]{14cm}
 supported by the German Federal Ministry for Education and Research (BMBF), under
 contract No. 05h09GUF, and the SFB 676 of the Deutsche Forschungsgemeinschaft (DFG) \
\end{minipage}\\
\makebox[3ex]{$^{ J}$}
\begin{minipage}[t]{14cm}
 supported by the Japanese Ministry of Education, Culture, Sports, Science and Technology
 (MEXT) and its grants for Scientific Research\
\end{minipage}\\
\makebox[3ex]{$^{ K}$}
\begin{minipage}[t]{14cm}
 supported by the Korean Ministry of Education and Korea Science and Engineering
 Foundation\
\end{minipage}\\
\makebox[3ex]{$^{ L}$}
\begin{minipage}[t]{14cm}
 supported by FNRS and its associated funds (IISN and FRIA) and by an Inter-University
 Attraction Poles Programme subsidised by the Belgian Federal Science Policy Office\
\end{minipage}\\
\makebox[3ex]{$^{ M}$}
\begin{minipage}[t]{14cm}
 supported by the Spanish Ministry of Education and Science through funds provided by
 CICYT\
\end{minipage}\\
\makebox[3ex]{$^{ N}$}
\begin{minipage}[t]{14cm}
 supported by the Natural Sciences and Engineering Research Council of Canada (NSERC) \
\end{minipage}\\
\makebox[3ex]{$^{ O}$}
\begin{minipage}[t]{14cm}
 partially supported by the German Federal Ministry for Education and Research (BMBF)\
\end{minipage}\\
\makebox[3ex]{$^{ P}$}
\begin{minipage}[t]{14cm}
 supported by RF Presidential grant N 4142.2010.2 for Leading Scientific Schools, by the
 Russian Ministry of Education and Science through its grant for Scientific Research on
 High Energy Physics and under contract No.02.740.11.0244 \
\end{minipage}\\
\makebox[3ex]{$^{ Q}$}
\begin{minipage}[t]{14cm}
 supported by the Netherlands Foundation for Research on Matter (FOM)\
\end{minipage}\\
\makebox[3ex]{$^{ R}$}
\begin{minipage}[t]{14cm}
 supported by the Israel Science Foundation\
\end{minipage}\\
\vspace{30em} \pagebreak[4]


\makebox[3ex]{$^{ a}$}
\begin{minipage}[t]{14cm}
now at University of Salerno, Italy\
\end{minipage}\\
\makebox[3ex]{$^{ b}$}
\begin{minipage}[t]{14cm}
now at Queen Mary University of London, United Kingdom\
\end{minipage}\\
\makebox[3ex]{$^{ c}$}
\begin{minipage}[t]{14cm}
also funded by Max Planck Institute for Physics, Munich, Germany\
\end{minipage}\\
\makebox[3ex]{$^{ d}$}
\begin{minipage}[t]{14cm}
also Senior Alexander von Humboldt Research Fellow at Hamburg University,
 Institute of Experimental Physics, Hamburg, Germany\
\end{minipage}\\
\makebox[3ex]{$^{ e}$}
\begin{minipage}[t]{14cm}
also at Cracow University of Technology, Faculty of Physics,
 Mathemathics and Applied Computer Science, Poland\
\end{minipage}\\
\makebox[3ex]{$^{ f}$}
\begin{minipage}[t]{14cm}
supported by the research grant No. 1 P03B 04529 (2005-2008)\
\end{minipage}\\
\makebox[3ex]{$^{ g}$}
\begin{minipage}[t]{14cm}
supported by the Polish National Science Centre, project No. DEC-2011/01/BST2/03643\
\end{minipage}\\
\makebox[3ex]{$^{ h}$}
\begin{minipage}[t]{14cm}
now at Rockefeller University, New York, NY
 10065, USA\
\end{minipage}\\
\makebox[3ex]{$^{ i}$}
\begin{minipage}[t]{14cm}
now at DESY group FS-CFEL-1\
\end{minipage}\\
\makebox[3ex]{$^{ j}$}
\begin{minipage}[t]{14cm}
now at Institute of High Energy Physics, Beijing, China\
\end{minipage}\\
\makebox[3ex]{$^{ k}$}
\begin{minipage}[t]{14cm}
now at DESY group FEB, Hamburg, Germany\
\end{minipage}\\
\makebox[3ex]{$^{ l}$}
\begin{minipage}[t]{14cm}
also at Moscow State University, Russia\
\end{minipage}\\
\makebox[3ex]{$^{ m}$}
\begin{minipage}[t]{14cm}
now at University of Liverpool, United Kingdom\
\end{minipage}\\
\makebox[3ex]{$^{ n}$}
\begin{minipage}[t]{14cm}
now at CERN, Geneva, Switzerland\
\end{minipage}\\
\makebox[3ex]{$^{ o}$}
\begin{minipage}[t]{14cm}
also affiliated with Universtiy College London, UK\
\end{minipage}\\
\makebox[3ex]{$^{ p}$}
\begin{minipage}[t]{14cm}
now at Goldman Sachs, London, UK\
\end{minipage}\\
\makebox[3ex]{$^{ q}$}
\begin{minipage}[t]{14cm}
also at Institute of Theoretical and Experimental Physics, Moscow, Russia\
\end{minipage}\\
\makebox[3ex]{$^{ r}$}
\begin{minipage}[t]{14cm}
also at FPACS, AGH-UST, Cracow, Poland\
\end{minipage}\\
\makebox[3ex]{$^{ s}$}
\begin{minipage}[t]{14cm}
partially supported by Warsaw University, Poland\
\end{minipage}\\
\makebox[3ex]{$^{ t}$}
\begin{minipage}[t]{14cm}
now at Istituto Nucleare di Fisica Nazionale (INFN), Pisa, Italy\
\end{minipage}\\
\makebox[3ex]{$^{ u}$}
\begin{minipage}[t]{14cm}
now at Haase Energie Technik AG, Neum\"unster, Germany\
\end{minipage}\\
\makebox[3ex]{$^{ v}$}
\begin{minipage}[t]{14cm}
now at Department of Physics, University of Bonn, Germany\
\end{minipage}\\
\makebox[3ex]{$^{ w}$}
\begin{minipage}[t]{14cm}
also affiliated with DESY, Germany\
\end{minipage}\\
\makebox[3ex]{$^{ x}$}
\begin{minipage}[t]{14cm}
also at University of Tokyo, Japan\
\end{minipage}\\
\makebox[3ex]{$^{ y}$}
\begin{minipage}[t]{14cm}
now at Kobe University, Japan\
\end{minipage}\\
\makebox[3ex]{$^{ z}$}
\begin{minipage}[t]{14cm}
supported by DESY, Germany\
\end{minipage}\\
\makebox[3ex]{$^{\dagger}$}
\begin{minipage}[t]{14cm}
 deceased \
\end{minipage}\\
\makebox[3ex]{$^{aa}$}
\begin{minipage}[t]{14cm}
member of National Technical University of Ukraine, Kyiv Polytechnic Institute,
 Kyiv, Ukraine\
\end{minipage}\\
\makebox[3ex]{$^{ab}$}
\begin{minipage}[t]{14cm}
member of National University of Kyiv - Mohyla Academy, Kyiv, Ukraine\
\end{minipage}\\
\makebox[3ex]{$^{ac}$}
\begin{minipage}[t]{14cm}
partly supported by the Russian Foundation for Basic Research, grant 11-02-91345-DFG\_a\
\end{minipage}\\
\makebox[3ex]{$^{ad}$}
\begin{minipage}[t]{14cm}
Alexander von Humboldt Professor; also at DESY and University of Oxford\
\end{minipage}\\
\makebox[3ex]{$^{ae}$}
\begin{minipage}[t]{14cm}
STFC Advanced Fellow\
\end{minipage}\\
\makebox[3ex]{$^{af}$}
\begin{minipage}[t]{14cm}
now at LNF, Frascati, Italy\
\end{minipage}\\
\makebox[3ex]{$^{ag}$}
\begin{minipage}[t]{14cm}
This material was based on work supported by the
 National Science Foundation, while working at the Foundation.\
\end{minipage}\\
\makebox[3ex]{$^{ah}$}
\begin{minipage}[t]{14cm}
also at Max Planck Institute for Physics, Munich, Germany, External Scientific Member\
\end{minipage}\\
\makebox[3ex]{$^{ai}$}
\begin{minipage}[t]{14cm}
now at Tokyo Metropolitan University, Japan\
\end{minipage}\\
\makebox[3ex]{$^{aj}$}
\begin{minipage}[t]{14cm}
now at Nihon Institute of Medical Science, Japan\
\end{minipage}\\
\makebox[3ex]{$^{ak}$}
\begin{minipage}[t]{14cm}
now at Osaka University, Osaka, Japan\
\end{minipage}\\
\makebox[3ex]{$^{al}$}
\begin{minipage}[t]{14cm}
also at \L\'{o}d\'{z} University, Poland\
\end{minipage}\\
\makebox[3ex]{$^{am}$}
\begin{minipage}[t]{14cm}
member of \L\'{o}d\'{z} University, Poland\
\end{minipage}\\
\makebox[3ex]{$^{an}$}
\begin{minipage}[t]{14cm}
now at Department of Physics, Stockholm University, Stockholm, Sweden\
\end{minipage}\\
\makebox[3ex]{$^{ao}$}
\begin{minipage}[t]{14cm}
also at Cardinal Stefan Wyszy\'nski University, Warsaw, Poland\
\end{minipage}\\

}

\newpage

\pagenumbering{arabic} 
\pagestyle{plain}

\section{Introduction}
\label{sec-int}

The study of deep inelastic scattering (DIS) of leptons off nucleons has
contributed significantly to tests of the Standard Model (SM) of the electroweak 
and strong interactions. The structure of nucleons has mainly 
been determined from DIS experiments. The $ep$ collider HERA has allowed an
 extension in the four-momentum-transfer squared, $Q^2$, and in Bjorken $x$ by several orders of magnitude with respect to previous
fixed-target experiments~\cite{rmp:71:1275}. The higher $Q^2$ reach of HERA has 
also allowed the exploration of the electroweak sector of the SM. 

The ZEUS and H1 collaborations have both measured the $e^-p$ and $e^+p$ neutral current (NC) DIS cross sections up to $Q^2$ of $30\,000\,\gev^2$ using the data collected in the years 1992--2000, referred to as the HERA\;I data-taking period. A combination of the results has been published~\cite{:2009wt}.

The combined cross sections were used as the sole input to a next-to-leading order (NLO) QCD analysis to determine the set of parton distribution functions (PDFs) called HERAPDF1.0~\cite{:2009wt}. The HERA\;I data were sufficiently precise to demonstrate the effects of $Z$ exchange by comparing the $e^-p$ and $e^+p$ NC DIS cross sections at high $Q^2$~\cite{:2009wt}.

HERA underwent a major upgrade before the 2003--2007 data-taking period, referred to as HERA\;II running.  The upgrade significantly increased the instantaneous luminosity delivered by
HERA and also provided longitudinally polarised electron\footnote{In this paper, the word ``electron'' refers to both electrons and positrons, unless otherwise stated.} beams for the collider experiments. The larger collected luminosity provided a higher reach in $Q^2$
and the longitudinal lepton-beam polarisation gave a unique opportunity to study the helicity structure of the electroweak interaction. 

The ZEUS collaboration has already published the NC and CC inclusive cross sections for all HERA\;II running periods except for the NC $e^+p$ data collected in 2006--2007~\cite{Chekanov:2006da, Chekanov:2008aa, Chekanov:2009gm, Collaboration:2010xc}. In this paper, we report NC $e^+p$ cross sections for $Q^2 > 185 \gev^2$ for this period. The H1 collaboration has recently also published NC and CC cross sections for the HERA\;II running periods~\cite{Aaron:2012qi}.

Parity-violating effects induced by electroweak processes can be demonstrated via the difference between the cross sections involving negatively and positively polarised electron beams.  For positrons, this is expressed through the asymmetry parameter $A^{+}$, which is proportional to the product of the electron axial ($a_e$)
and quark vector ($v_q$) couplings to the $Z$ boson. 
In this paper, the cross sections and the polarisation asymmetry are presented and compared to
SM predictions, providing a test of the electroweak sector and a key input to further QCD fits.
\section{Predictions from the Standard Model}\label{sec-pred}

Inclusive deep inelastic lepton-proton scattering can be described in
terms of the kinematic variables $x$, $y$, and $Q^2$.  The variable
$Q^2$ is defined as $Q^2 = -q^2 = -(k-k')^2$, where $k$ and $k'$ are
the four-momenta of the incoming and scattered lepton, respectively.
Bjorken $x$ is defined as $x=Q^2/2P \cdot q$, where $P$ is
the four-momentum of the incoming proton. The fraction of the lepton energy transferred to the proton in the rest frame
of the proton is given by $y = P \cdot q / P \cdot k$. The variables $x$, $y$ 
and $Q^2$ are related by $Q^2=sxy$, where  $s$ is the square of the lepton-proton centre-of-mass energy. At HERA, $s=4E_e E_p$, where $E_{e}$ and $E_{p}$ 
are the initial electron and proton energies, respectively. 
The electroweak Born-level cross section for $e^ \pm p$ NC
interactions can be written as~\cite{devenish:2003:dis, zfp:c24:151}
\begin{equation}
  \ddif{\sigma(e^{\pm}p)}{x}{Q^{2}} = 
  \frac{2 \pi \alpha^{2} }{xQ^{4}}
  [Y_{+} \tilde{F_{2}}(x,Q^{2})  
  \mp Y_{-} x\tilde{F_{3}}(x,Q^{2})  
  - y^{2}\tilde {F_{L}}(x,Q^{2})],
\label{eqn:unpol_xsec}
\end{equation}
where $\alpha$ is the fine-structure constant,
$Y_{\pm} = 1 \pm (1 - y)^{2}$ and
$\tilde{F_{2}}(x,Q^{2})$, $\tilde{F_{3}}(x,Q^{2})$ and
$\tilde{F_{L}}(x,Q^{2})$
are generalised structure functions. NLO
 QCD calculations predict~\cite{devenish:2003:dis, zfp:c24:151} and measurements confirm~\cite{Chekanov:2009na, Collaboration:2010ry} that the contribution of the longitudinal structure 
function, $\tilde {F_L}$, to $d^2\sigma /dx dQ^2$ is approximately $1\%$, averaged 
over the kinematic range considered here. 
However, in the  high-$y$ region, the $\tilde {F_L}$ contribution to 
the cross section can be as large as $10\%$ and it is therefore
 included in the SM predictions  compared to the measurements 
presented in this paper.  

The generalised structure functions depend on the longitudinal polarisation 
of the lepton beam, which is defined as
\begin{equation}
P_{e}=\frac{N_{R}-N_{L}}{N_{R}+N_{L}}, 
\label{eqn:pol}
\end{equation}
where $N_{R}$ and $N_{L}$ are the numbers of right- and left-handed leptons in
the beam\footnote{At the HERA beam energies, the mass of the incoming leptons can be neglected,
and therefore the difference between handedness and helicity can also be neglected.}.

The $\tilde{F_{2}}$ term in Eq.~(\ref{eqn:unpol_xsec}) is dominant at low $Q^{2}$, where only photon exchange is important, 
while the $\tilde{F_{3}}$ term starts to contribute significantly to the cross section only at  
$Q^{2}$ values of the order of the mass of the $Z$ boson squared, $M_Z^2$, and above. It results from $\gamma / Z$
interference and $Z$ exchange. 
The sign of the $\tilde{F_{3}}$ term in Eq.~(\ref{eqn:unpol_xsec}) shows that
electroweak effects decrease (increase) the $e^{+} p$ ($e^{-} p$) cross section. 

The reduced cross sections for $e^-p$ and $e^+p$ scattering are defined as
%
\begin{equation}
  \tilde{\sigma}^{e^{\pm} p} 
  =
  \frac {xQ^{4}} {2 \pi \alpha^{2} }
  \frac {1} {Y_{+}}
  \ddif{\sigma(e^{\pm}p)}{x}{Q^{2}}
  =
  \tilde{F_{2}}(x,Q^{2}) \mp \frac {Y_{-}} {Y_{+}} x \tilde{F_{3}}(x,Q^{2})- \frac {y^2} {Y_{+}} F_{L}(x,Q^{2}).
\label{eqn:red}
\end{equation}
Thus $x\tilde{F_3}$ can be obtained from the difference of the $e^{-} p$ and $e^{+} p$ 
reduced cross sections as
%
%
\begin{equation}
 x\tilde{F_3}
  =  \frac {Y_{+}} {2Y_{-}}( \tilde{\sigma}^{e^{-} p} - \tilde{\sigma}^{e^{+} p} ).
\label{eqn:xf3}
\end{equation} 
  
The generalised structure functions can be split into terms depending on $\gamma$
exchange ($F_2^{\gamma}$), $Z$ exchange ($F_2^Z$, $xF_3^Z$) and $\gamma/Z$ interference
($F_2^{\gamma Z}$, $xF_3^{\gamma Z}$) as
\begin{equation}
    \tilde{F_2}^\pm = F_2^{\gamma} - (v_e \pm P_e a_e) \chi_{Z} F_2^{\gamma Z} +
  (v_e^2 + a_e^2 \pm 2 P_e v_e a_e) {\chi_{Z}^{2}} F_2^{Z}  ,
\label{eqn:gen_f2}
\end{equation}
\begin{equation}
   x\tilde{F_3}^\pm =  - (a_e \pm P_e v_e) \chi_{Z} xF_3^{\gamma Z} + (2 v_e a_e 
\pm P_e(v_e^2 + a_e^2)) {\chi_{Z}^{2}} xF_3^{Z}.
\label{eqn:gen_xf3}
\end{equation}
The SM predictions for the respective vector and axial couplings of the electron to the $Z$ boson are $v_{e} = -1/2 + 2\sin^2\theta_W$ and $a_{e} = -1/2$, where $\theta_W$ is the
Weinberg angle. The relative fraction of events coming from $Z$ with respect to $\gamma$ exchange is given by 
\begin{equation}
\chi_{Z}=\frac{1}{\sin^2{2\theta_W}} \frac{Q^{2}}{M_{Z}^{2}+Q^{2}} .
  \label{eqn:chiz}
\end{equation}
This fraction varies between 0.03 and 1.1 over the range of the analysis, $185~\Gev^2 < Q^2 < 50\,000~\Gev^2$. 
For the unpolarised case ($P_{e} = 0$), ignoring terms containing $v_e$, which is small ($\approx -0.04$), the interference structure 
function, $xF_3^{\gamma Z}$, is the dominant term in $x\tilde{F_3}$, and 
\begin{equation}
  x\tilde{F_3} \simeq  - a_e \chi_{Z} xF_3^{\gamma Z}. 
\label{eqn:gen_xf3gz}
\end{equation}
   




The structure functions can be written in terms of the sum and differences of the quark and anti-quark momentum
distributions. At leading order (LO) in QCD,
\begin{equation}
[F_2^{\gamma},F_2^{\gamma Z},F_2^{Z}] = 
\sum _q [e_{q}^{2}, 2e_{q}v_{q},v_{q}^{2}+a_{q}^{2}] 
 x (q + \bar{q}),
\label{eqn:struc1}
\end{equation}
\begin{equation}
[xF_3^{\gamma Z},xF_3^{Z}] = 
\sum _q [e_{q}a_{q},v_{q}a_{q}] 
 2x (q - \bar{q}),
\label{eqn:struc2}
\end{equation}

where $v_{q}$ and $a_{q}$ are the respective vector and axial couplings of the quark $q$ to
the $Z$ boson, and $e_{q}$ is the electric charge of the quark. 
The densities of the quarks and anti-quarks are given by $q$ and $\bar{q}$,
respectively. The sum runs over all quark flavours except the top quark.
 
The sensitivity of $xF_3^{\gamma Z}$ to the $u$ and $d$ valence-quark momentum 
distributions is demonstrated in LO QCD through the expression
\begin{equation}
 xF_3^{\gamma Z} 
  =  2x [e_u a_u u_v + e_d a_d d_v] = \frac{x}{3}(2u_{v} + d_{v}),
\label{eqn:xf3gz_simple}
\end{equation} 
where the SM values $v_{u} = 1/2 -4/3\sin^2\theta_W$ and $a_{u} = 1/2$ have been used.%
%


%

The charge-dependent polarisation asymmetry, $A^+$, for a pure right-handed
($P_{e} = +1$) and left-handed ($P_{e} = -1$) positron beam is defined
as
\begin{equation}
A^+ \equiv \frac{\sigma^{+}(P_{e} = +1) -\sigma^{+}(P_{e} = -1)}{\sigma^{+}(P_{e} =  +1) +\sigma^{+}(P_{e} = -1)}, 
\label{eqn:asymDefn}
\end{equation} 
where $\sigma^{+}(P_{e}=+1)$ and $\sigma^{+}(P_{e}=-1)$ are the
differential $e^+p$ cross sections 
evaluated at longitudinal polarisation values of $+1$ and $-1$, respectively. 
In general, $A^+$ can be calculated as

\begin{equation}
A^{+} = \frac{\sigma^+(P_{e,+}) - \sigma^+(P_{e,-})}
{P_{e,+}\sigma^+(P_{e,-}) - P_{e,-}\sigma^+(P_{e,+})} \ ,
  \label{eqn:asymMeas}
\end{equation}

where $\sigma^{+}(P_{e,+})$ and $\sigma^{+}(P_{e,-})$ are the
differential $e^+p$ cross sections evaluated at any positive and negative 
polarisation values. 
For $P_{e,+}\approx -P_{e,-}$ this equation becomes 
\begin{equation}
  A^+ =\frac{2}{P_{e,+}- P_{e,-}} \cdot \frac{\sigma^+(P_{e,+}) - \sigma^+(P_{e,-})} { \sigma^+(P_{e,+}) +\sigma^+(P_{e,-})}.
  \label{eqn:asymapp}
\end{equation}
%
Keeping only the leading terms, $A^+$ can be written as
\begin{equation}                               
  A^+ \simeq  -\chi_{Z} a_e \frac{F_2^{\gamma Z}}{F_2^{\gamma}} 
= -2 \chi_{Z}a_e v_q e_q/e_q^2 \propto a_e v_q. 
\label{fgf}   
\end{equation}
As the asymmetry parameter is proportional to the ratio of the $F_2^{\gamma Z}$ and $F_2^{\gamma}$ structure
functions, it is to first order insensitive to PDFs. 
Therefore a measurement of $A^+$ can give direct evidence of parity violation with minimal assumptions on the proton structure. As, in the SM, $A^+$ is expected to be a small quantity, less than 10\% for $Q^2$ values below $2\,000\,\gev^2$, increasing slowly to 30\% by $Q^2$ of $10\,000\,\gev^2$, precise measurements of the polarised cross sections are required.

\section{Experimental set-up}

\label{sec-ncdet}

The analysis is based on a data sample collected in 2006--2007, when HERA collided positrons of energy $E_e = 27.5 \gev$ with protons of energy $E_p = 920\gev$, corresponding to a centre-of-mass energy $\sqrt{s} = 318 \gev$. The total integrated luminosity of the sample is $135.5 \pm 2.5 \pbi$, of which $78.8 \pm 1.4 \pbi$ were collected at a luminosity-weighted lepton-beam polarisation $P_e = 0.32 \pm 0.01$ and $56.7 \pm 1.1 \pbi$ at $P_e = -0.36 \pm 0.01$.  

\Zdetdesc

Charged particles were tracked in the central tracking detector (CTD)~\citeCTD which operated in a magnetic field of $1.43~{\rm T}$ provided by a thin superconducting solenoid. The CTD consisted of 72 cylindrical drift chamber layers, organised in nine superlayers covering the polar-angle\footnote{The ZEUS coordinate system is a right-handed Cartesian system, with the $Z$ axis pointing in the
proton beam direction, referred to as the ``forward direction'', and the $X$ axis pointing towards the centre of HERA. The coordinate origin is at the nominal interaction point. The pseudorapidity is
defined as  $\eta = - \ln(\tan (\theta/2))$, where the polar angle, $\theta$, is measured with respect to the proton beam
direction.} region $15^\circ < \theta  < 164^\circ$. The CTD was complemented by a silicon microvertex detector (MVD)~\citeMVD, consisting of three active layers in the barrel and four disks in the forward region. For CTD-MVD tracks that pass through all nine CTD
superlayers, the momentum resolution was $\sigma(p_T )/p_T = 0.0029 p_T \oplus 0.0081 \oplus 0.0012/p_T$, with $p_T$ in GeV.

\Zcaldesc



\Zlumidesc{1.8~\% for the period with $P_e = 0.32$ and 1.9~\% for the period with $P_e = -0.36$}

The lepton beam in HERA became naturally transversely polarised
through the Sokolov-Ternov effect~\cite{sovpdo:8:1203}.
The characteristic build-up time in HERA
was approximately 40 minutes.
Spin rotators on either side of the ZEUS detector
changed the transverse polarisation of the beam
into longitudinal polarisation and back to transverse.
The electron beam polarisation was measured
using two independent polarimeters,
the transverse polarimeter (TPOL)~\cite {Baier:1969hw,nim:a329:79} and
the longitudinal polarimeter (LPOL)~\cite {nim:a479:334}.
Both devices exploited the spin-dependent cross section
for Compton scattering of circularly polarised photons off electrons
to measure the beam polarisation.
The luminosity and polarisation measurements were made over time scales
that were much shorter than the polarisation build-up time.

\section{Monte Carlo simulation}
\label{sec-mc}

Monte Carlo (MC) simulations were used to determine the efficiency of 
the event selection, the accuracy of the kinematic reconstruction, to estimate 
the background rate and to extrapolate the measured cross sections to the full 
kinematic region. The effective luminosities of the MC samples were at least five times 
larger than that of the data sample and were normalised to the total integrated luminosity of the data.

Neutral current DIS events were simulated, including radiative effects, using the
{\sc Heracles}~\cite{cpc:69:155} program with the {\sc Djangoh}
1.6~\cite{proc:hera:1991:1419,*spi:www:djangoh11}  interface to the
hadronisation programs and using the CTEQ5D \cite{epj:c12:375} PDFs.
The hadronic final state was simulated using the 
colour-dipole model of {\sc Ariadne} 4.12~\cite{cpc:71:15}.
To investigate systematic uncertainties, the {\sc Meps} model of {\sc Lepto}~6.5~\cite{pl:b366:371} was also used.
The Lund string model of {\sc Jetset} 
7.4~\cite{cpc:39:347,*cpc:43:367,*cpc:82:74} was used for the hadronisation. 
Photoproduction ($\gamma p$) events
were simulated using
{\sc Herwig}~5.9~\cite{cpc:67:465} to study this background. 

The ZEUS detector response was simulated using a program based on {\sc Geant}
3.21~\cite{tech:cern-dd-ee-84-1}. The generated events were passed through the
detector simulation, subjected to the same trigger requirements as the data
and processed by the same reconstruction programs. 

The distribution of the Z position of the interactions was
a crucial input to the MC simulation with which the 
event-selection efficiency was determined. In order to
measure this distribution, a special NC DIS sample was
selected, for which the event selection efficiency did
not depend on the Z of the interaction\cite{stewart:phd:2012}.


\section{Event reconstruction}
\label{sec-recon}

Neutral current events at high $Q^2$ are characterised by the presence of an
isolated high-energy electron in the final state. The transverse momentum of 
the scattered electron balances that of the hadronic final state. Therefore
the net transverse momentum of the event, $P_T$, representing the vectorial sum of the transverse momenta of all particles, $\vec{p}_T$, should be small. 
The measured net $P_T$ and transverse energy, $E_T$, were calculated as
\begin{alignat}{2}
  P_T^2 & = & P_X^2 + P_Y^2 = & \left( \sum\limits_{i} E_i \sin \theta_i \cos
    \phi_i \right)^2+ \left( \sum\limits_{i} E_i \sin \theta_i \sin \phi_i
  \right)^2,
  \label{eq-PT2}\\ 
  E_T & = & \sum\limits_{i} E_i \sin \theta_i, \nonumber
\end{alignat}
where the sum ran over all calorimeter energy deposits, $E_i$, and the
polar ($\theta_i$) and azimuthal ($\phi_i$) angles were measured with respect to the interaction vertex.
The variable $\delta$, defined as
\begin{equation}
  \delta \equiv \sum\limits_{i} (E-P_Z)_{i} = \sum\limits_{i} ( E_i - E_i \cos
  \theta_{i} )\equiv E-P_Z,
  \label{eq-Delta}
\end{equation}
was also used in the event selection. 
Conservation of energy and longitudinal momentum implies that  
$\delta = 2E_e= 55 \gev$, if all final-state particles were
detected and perfectly measured. Undetected particles that escape through the
forward beam-hole had a negligible effect on $\delta$. However, particles 
lost through the rear beam-hole could lead to a substantial reduction in
$\delta$. This was the case for $\gamma p$ events, where the electron
emerged at a very small scattering angle, or for events in which an
initial-state bremsstrahlung photon was emitted. 

The CAL energy deposits were separated into those associated 
with the scattered electron and all other energy deposits. The sum of
the latter was called the hadronic energy. 
The hadronic polar angle, $\gamma_h$, was defined as
\begin{equation}
  \cos\gamma_h = \frac{P_{T,h}^2 - \delta^2_h}{P_{T,h}^2 + \delta^2_h},
  \label{eqn:gamma_h}
\end{equation}
where the quantities $P_{T,h}$ and $\delta_h$ were derived from
Eqs.~(\ref{eq-PT2}$-$\ref{eq-Delta}) using only the hadronic energy. In the na\"{\i}ve quark-parton model, $\gamma_h$ is the angle by which the struck quark is scattered.

The double angle (DA) method~\cite{proc:hera:1991:23,*proc:hera:1991:43}
used the polar angle of the scattered electron, $\theta_e$, and the
hadronic angle, $\gamma_h$, to reconstruct the kinematic variables
$x_\DA$, $y_\DA$, and $Q^2_\DA$. For the determination of $\theta_e$, 
tracking information was also used when available. 
The DA method was insensitive to uncertainties
in the overall energy scale of the calorimeter. However, it was sensitive to
initial-state QED radiation and an accurate simulation of 
the detector response was necessary.
The variable $y$ was reconstructed using the electron
method ($y_e$). The Jacquet-Blondel method 
($y_\JB$) \cite{proc:epfacility:1979:391} was used in the event
selection in kinematic regions where it 
provided better resolution.
 

\section{Event selection}
\label{sec-sel}

\subsection{Trigger requirements}
\label{sec-Trigger}
Events were selected using a three-level trigger
system~\cite{zeus:1993:bluebook,uproc:chep:1992:222,nim:a580:1257}.  
At the first level, only
coarse calorimeter and tracking information was available. 
Events were selected if they had an energy deposit in the CAL
consistent with an isolated electron. In addition, events with high $E_{T}$ or high energy in the electromagnetic part of the calorimeter in
coincidence with a CTD track were selected. At the second level, a requirement
on $\delta$ was used to select NC DIS events. Timing information from the
calorimeter was used to reject events inconsistent with the bunch-crossing
time. At the third level, events were fully reconstructed. The requirements were similar to, but looser than the offline cuts
described below.

 
\subsection{Offline requirements}
\label{sec-ncsel}
The following criteria were imposed to select NC events.
\begin{itemize}
\item {Electron identification:}\\ 
  an algorithm which combined information from the energy deposits 
  in the calorimeter with tracks measured in 
  the central tracking detectors was used to identify the scattered 
  electron~\cite{epj:c11:427}. To ensure a high purity 
  and to reject background, the identified electron was required to have 
  an energy, $E_e^\prime$, of at least $10 \gev$ and to be isolated such that the energy not 
  associated with the electron in an $\eta-\phi$ cone of radius 0.8 centred on
  the electron was less than $5 \gev$.\\ 
  A track matched to the energy deposit in the calorimeter was required for
  events in which an electron was found within the region of good acceptance of the tracking
  detectors, which was $0.3<\theta<2.5$~\cite{januschek:phd:2011}. 
  This was done by restricting the distance of
  closest approach (DCA) between the track extrapolated to the calorimeter
  surface and the energy cluster position to within 10 cm and by requiring an 
  electron track momentum ($p_{e}^{\rm trk}$) larger than $3 \gev$. 


  A matched track was not required if the electron emerged outside the acceptance of the tracking detectors.

\item {Background rejection:}\\  
the requirement $38 < \delta < 65 \gev$ was imposed to remove 
$\gamma p$ and beam-gas events and to reduce the number of events with significant 
QED initial-state radiation. To further reduce background from $\gamma p$ events,
$y_{e}$ was required to be less than $0.9$. 
The measured $P_T$ was expected to be small for NC events. Therefore, 
in order to remove cosmic rays  and beam-related background events, 
the quantity $P_{T}/\sqrt{E_{T}}$ was required to be less than
$4\sqrt{\gev}$,  
and the quantity $P_{T}/E_{T}$ was required to be less than 0.7.

\item {Additional requirements:}\\ 
the projection of $\gamma_h$ onto the face of FCAL was required to be outside a
radius of $18~{\mathrm{cm}}$ centred on the beam-pipe axis, to reject events
where most of the hadronic final state was lost in the forward beam-pipe.

The $Z$ coordinate of the $ep$ interaction vertex, reconstructed using 
tracks in the CTD and the MVD, was required to satisfy $| Z_{\rm vtx} | < 30$~cm.

In order to avoid the kinematic region in which the MC simulation is not appropriate due to missing higher-order QED corrections~\cite{proc:hera:1991:1419,*spi:www:djangoh11}, the requirement $y_{\rm JB}(1-x_{\rm DA})^2>0.004$ was applied.

The final event sample was selected by requiring 
$Q^{2}_{\rm DA}>185 \gev^2$.
\end{itemize}

A total of 302,073 candidate events passed the selection criteria. 
The background contamination estimated from the $\gamma p$ MC was
about 0.2\% overall. 

\begin{figure}[p]
\vfill
\begin{center}
\includegraphics[width=6in]{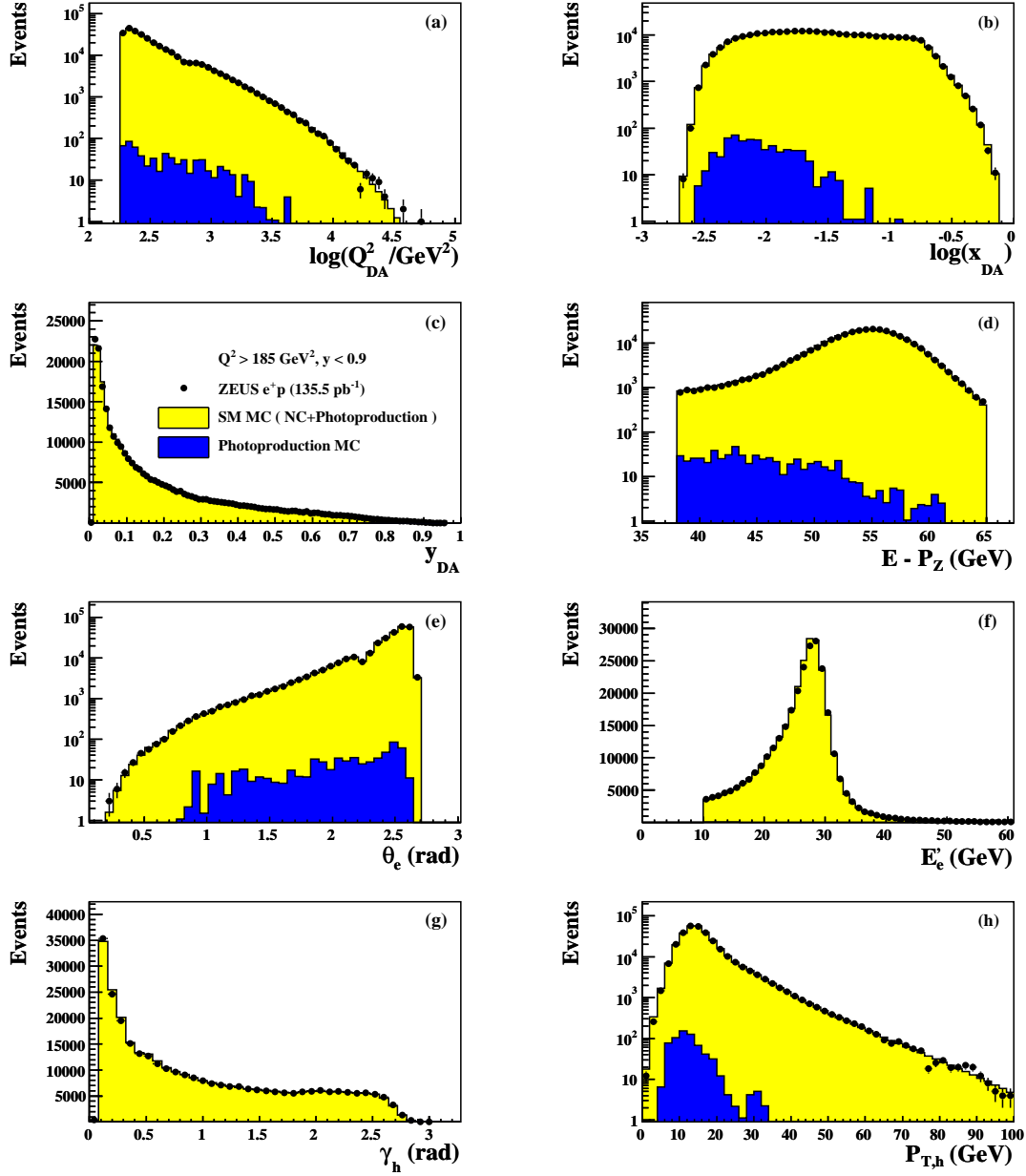}
\end{center}
\caption{
  Comparison of the $e^+ p$ NC data sample with the predictions from 
  the MC simulation.
  The MC distributions are normalised to the data luminosity.
  The distributions of 
  (a) $Q^{2}_{\rm DA}$, 
  (b) $x_{\rm DA}$, 
  (c) $y_{\rm DA}$, 
    (d) $E-P_{\rm Z}$,
  (e) $\theta_{\rm e}$,
  (f) $E_{\rm e}^{\prime}$, 
  (g) $\gamma_{h}$
  and 
  (h) $P_{T,h}$ 
  are shown.
}
\label{fig-cont}
\vfill
\end{figure}

Figure~\ref{fig-cont} shows a comparison between 
data and MC distributions for the variables $Q^{2}_{\rm DA}$, $x_{\rm DA}$, $y_{\rm DA}$, $E -P_Z$ of the event, $\theta_e$ and $E_e^\prime$ of the scattered electron and $\gamma_h$ and
$P_{T,h}$ of the final hadronic system. The description of the data by the MC is good.
\section{Cross-section determination}
\label{sec-xsecdet}
The kinematic region of the measurement was defined as  
$Q^2 > 185\gev^2$, $y < 0.9$ and $y(1-x)^2 > 0.004$. 
The single-differential cross-sections
$d\sigma/dQ^2$, $d\sigma/dx$ and $d\sigma/dy$
and the reduced cross-section $\tilde{\sigma}^{e^+p}$
were measured. In addition, the single-differential cross-sections
$d\sigma/dx$ and $d\sigma/dy$ were measured 
for the restricted range $Q^2 > 3\,000\gev^2$, $y< 0.9$ and $y(1-x)^2 > 0.004$.
The cross section in a particular bin
($d^2\sigma/dx dQ^2$ is used as an example) was determined according to
\begin{equation}
  \frac{d^2 \sigma}{dx dQ^2} = \frac{N_{\rm data}-N_{\rm bg}}{N_{\rm MC}}
  \cdot \frac{d^2 \sigma^{\rm SM}_{\rm Born}}{dx dQ^2} \ ,
\label{eq-xsect}
\end{equation}
where $N_{\rm data}$ is the number of data events in the bin,
$N_{\rm bg}$ is the number of background events
predicted from the photoproduction MC and
$N_{\rm MC}$ is the number of signal MC events
normalised to the luminosity of the data.
The SM prediction for the Born-level cross section,
$d^2 \sigma^{\rm SM}_{\rm Born}/dx dQ^2$, was evaluated
using the CTEQ5D PDFs~\cite{epj:c12:375} as used for the MC simulation
and using the PDG~\cite{epj:c15:1} values 
for the fine-structure constant, the mass of the $Z$ boson
and the weak mixing angle.
This procedure implicitly takes into account the acceptance,
bin-centering, and radiative corrections from the MC simulation.
The bin sizes used for the determination of the single-differential and reduced
cross sections were chosen to be commensurate with the detector resolutions.
The statistical uncertainties on the cross sections
were calculated from the number of events observed in the bins,
taking into account the statistical uncertainty of the MC simulation
(signal and background).
Poisson statistics were used for all bins.
\section{Systematic uncertainties}
\label{sec-sys}
Systematic uncertainties were estimated~\cite{stewart:phd:2012, januschek:phd:2011} by re-calculating the cross
sections after modifying the analysis, in turn, for the uncertainties detailed below.

\begin{itemize}

\item $\delta_1$: the variation of the electron energy scale in the MC 
by its estimated uncertainty of $\pm 1 \%$ resulted in changes of less than $0.5\%$
in the cross sections over most of the kinematic region,
due to the use of the DA reconstruction method.
The effect was at most  $3\%$ in the high-$y$ region of $d\sigma/dy$;

\item $\delta_2$: the uncertainties due to ``overlay'' events,
in which a DIS event overlapped with additional energy deposits
from some other interaction in the RCAL,
were estimated by narrowing or widening the $38 < \delta < 65\gev$ interval 
symmetrically by $\pm 2\gev$.\footnote{This would also affect remaining
photoproduction events. However, their contribution was negligable.} 
The effect on the cross sections was typically
below 1\%. In a few high-$Q^2$ bins, the uncertainty was
as large as 5\%, reaching 11\% in one reduced-cross-section bin;

\item $\delta_3$: systematic uncertainties
arising from the normalisation of the photoproduction background
were determined by changing the background normalisation
by its estimated uncertainty of $\pm 50\%$\cite{wlasenko:phd:2009}.
The resulting changes in the cross sections were typically below $0.5\%$,
reaching about $2\%$ in the medium-$Q^2$ reduced-cross-section bins;

\item $\delta_4$: to estimate the systematic uncertainty associated with the electron finder, 
an alternative electron-finding algorithm~\cite{nim:a365:508}
was used and the results were compared to those obtained using the nominal algorithm. The  systematic uncertainty from the electron-finding procedure was below 1\% for most of the phase space;

\item $\delta_5$: the upper limit of the $\theta$ range for which a matched track for the electron candidate was required was varied by  $\pm 0.1$ to account for uncertainties in the track-matching efficiency towards the edge of the CTD and BCAL. The uncertainty was mostly below $1.0 \%$, but about $2\%$ for the lower-$Q^2$ region;

\item $\delta_6$: the systematic uncertainty due to the choice of the parton-shower scheme
was evaluated by using the {\sc Meps} model of {\sc Lepto}
to calculate the acceptance instead of 
{\sc Ariadne}\footnote{Since the simulation of the parton showers could, 
in principle, also have an influence on the electron isolation,
the comparison was made removing the requirements on  
the electron isolation in order 
to prevent double counting of systematic uncertainty.
However, no measurable influence of the isolation cut on 
$\delta_6$ was observed.}.
The uncertainty was typically within $3\%$, but reached up to 8\%
in some bins of the reduced cross section and 
the highest bins of $d\sigma/dy$; 

\item $\delta_7$: the simulation of the first-level trigger was corrected in order to match the
measured efficiency in the data. The systematic effect of the uncertainty of the correction on the cross section was 
typically less than $1\%$, but reached about $2\%$ for medium $Q^2$ and high $y$;

\item $\delta_8$: to evaluate the systematic uncertainty 
related to the electron isolation criterion, the
isolation requirement was changed
by $\pm 2 \gev$ from its nominal value of $5 \gev$. 
The cross sections typically changed by much less than $0.5\%$;

\item $\delta_{9}$: the DCA requirement was changed from $10$ to $8 \cm$
to estimate the uncertainty in the background contamination
due to falsely identified electrons.
The uncertainties in the cross sections associated with
this variation were below $1\%$ over most of the kinematic range;

\item $\delta_{10}$: the energy resolution used in the MC
for the scattered electron was varied by its estimated uncertainty $\pm 1\%$.
The effect on the cross sections was mostly less than $0.5 \%$ and less than $1 \%$ over the full kinematic range;

\item $\delta_{11}$: to account for differences of the $p_e^{\rm trk}$
distributions in data and MC,
the $p_e^{\rm trk}$ requirement was varied by $\pm 1 \gev$,
resulting in a variation of the cross section by less than $0.5\%$
over most of the kinematic range,
and up to $6\%$ in a few reduced-cross-section bins;

\item $\delta_{12}$: the cut of $18 \cm$ on the projected radius
of the hadronic angle onto the FCAL was varied by $\pm 2 \cm$.
The cross sections typically changed by much less than $0.5\%$.
The effect rises up to a maximum of $7\%$ for the highest bins
of both $d\sigma/dy$ and the reduced cross section;

\item $\delta_{13}$: the variation of the hadronic energy scale by its estimated uncertainty of $\pm2\%$ in the MC resulted in changes of  mostly below 0.5\% and always less than 2\% in the cross sections over the full  kinematic range;

\item $\delta_{14}$: the systematic uncertainty associated
with cosmic-ray rejection was evaluated
by varying the $P_T/\sqrt{E_T}$ cut by $\pm 1 \sqrt{\Gev}$ and the $P_T/E_T$ cut by $\pm 0.1$ .
The cross-section uncertainties were mostly below $0.5 \%$
reaching a maximum of $6\%$ in one reduced-cross-section bin for the variation of the $p_T/\sqrt{E_T}$ cut;

\item $\delta_{15}$: The limit on the accepted $|Z_{\rm vtx}|$ was varied by  $\pm 5 \cm$, resulting in less than a $1\%$ change in the cross sections over most of the kinematic range, reaching a maximum of $6\%$ in the highest-$Q^2$ bins.

\end{itemize}
 
The 15 sources of systematic uncertainty were treated as
uncorrelated to each other. Bin-to-bin correlations
were found for $\delta$ 1,2,3,4,6,8,10,12 and 13.  
The positive and negative deviations from the nominal cross-section values
were added in quadrature separately
to obtain the total positive and negative systematic uncertainty.

The relative uncertainty in the measured polarisation was
4\%.  This has a negligible effect on the cross sections. 
The choice of which polarimeter to consider was made run-by-run
to maximise the available luminosity
and minimise the uncertainty in the measured polarisation.

The measured luminosity had a relative uncertainty of $1.8\%$ for the period with right-handed and $1.9\%$ for the period with left-handed polarisation.
The uncertainties in the luminosity and polarisation measurements were 
not included in the total systematic uncertainty shown in the final results.
\section{Results}
\label{sec-res}

\subsection{Unpolarised cross sections} 
\begin{figure}[p]
\vfill
\begin{center}
\includegraphics[width=1.0\textwidth]{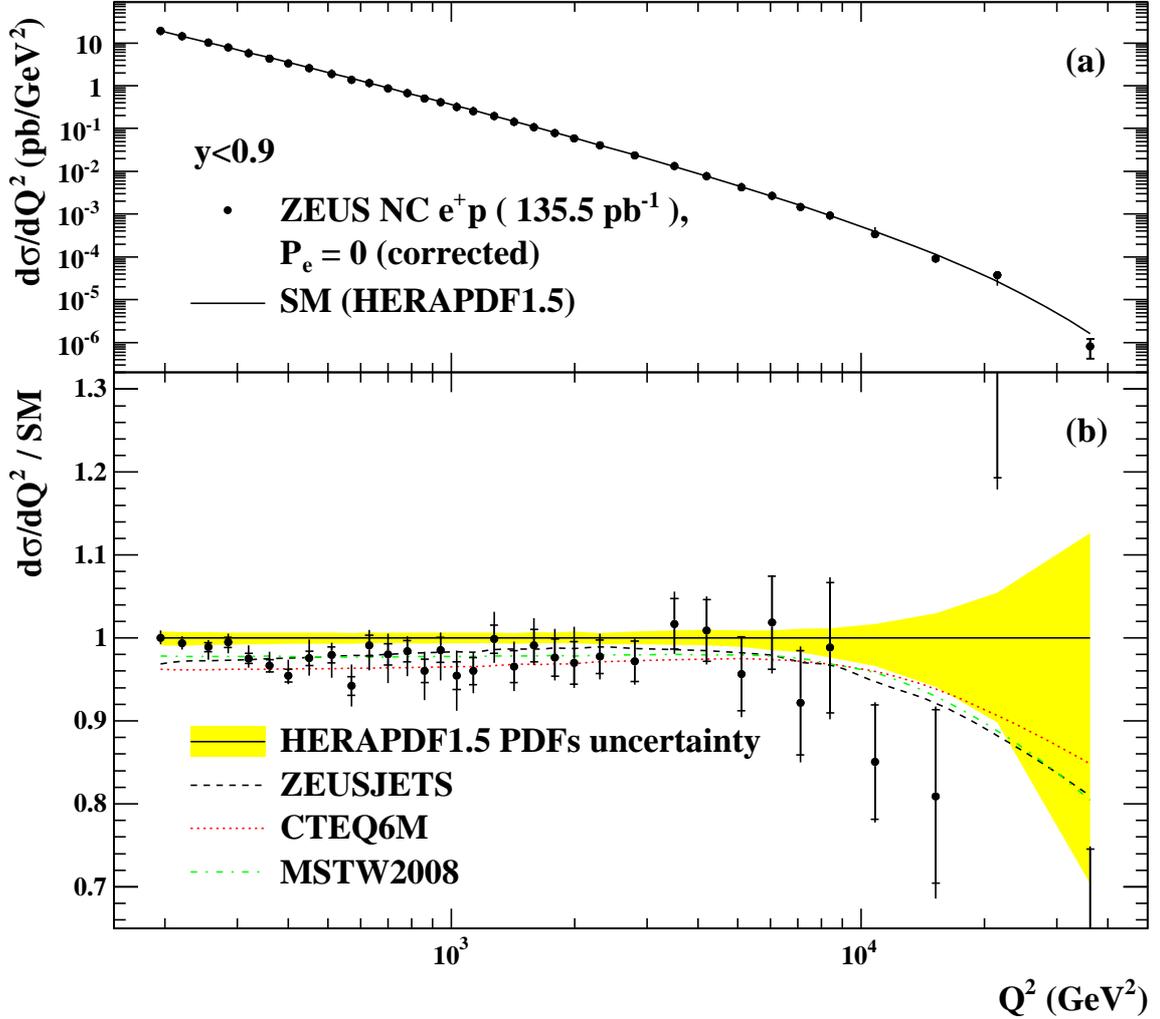}
\end{center}
\caption{
  (a) The $e^+ p$ NC DIS cross-section $d\sigma/dQ^{2}$ for $y < 0.9$ 
  and $y(1-x)^2>0.004$
  corrected to $P_e=0$ and (b) the ratio to the SM prediction.
  The closed circles represent data points in which
  the inner error bars show the statistical uncertainty
  while the outer bars show the statistical and systematic
  uncertainties added in quadrature.
  The curves show the predictions of the SM evaluated using
  the HERAPDF1.5 PDFs 
  and the shaded band shows the uncertainties from the HERAPDF1.5 PDFs. In the ratio plot in addition the ratios between other PDFs (ZEUSJETS (dashed), CTEQ6M (dotted) and MSTW2008 (dash-dotted)) and HERAPDF1.5 are shown as curves. The uncertainties of CTEQ6M and MSTW2008 are of the same order as of HERAPDF1.5, the uncertainties of ZEUSJETS are about a factor 2 higher.
}
\label{fig-q2sing}
\vfill
\end{figure}

\begin{figure}[p]
\vfill
\begin{center}           
            \includegraphics[width=1.0\textwidth]{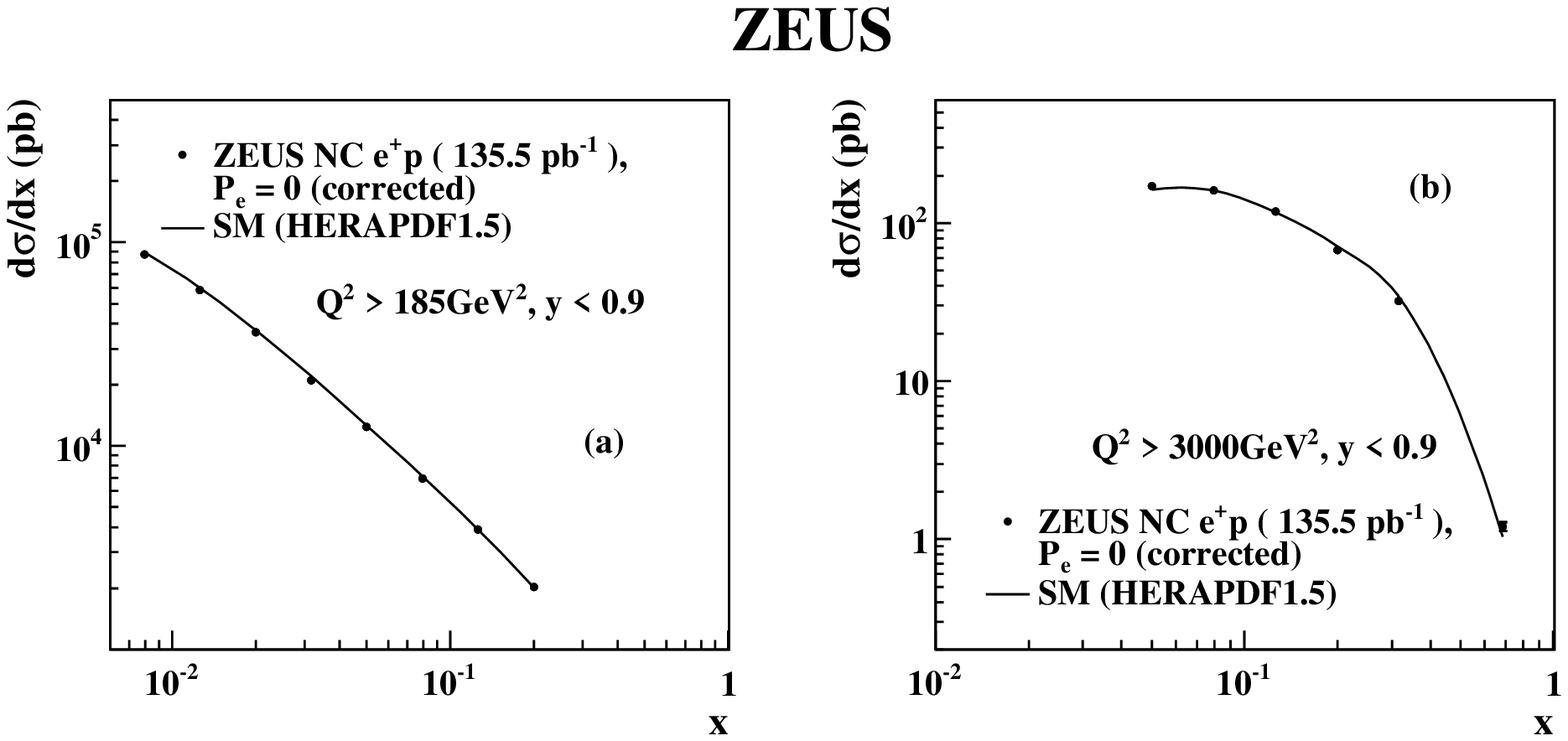}
\end{center}
\caption{
  The $e^+ p$ NC DIS cross-section $d\sigma/dx$ for (a) $Q^2 > 185 \gev^2$ 
  and (b) $Q^2 > 3\;000 \gev^2$
  for $y < 0.9$ and $y(1-x)^2>0.004$.
  The closed circles represent data points in which
  the inner error bars show the statistical uncertainty
  while the outer bars show the statistical and systematic
  uncertainties added in quadrature.
  The curves show the predictions of the SM evaluated using
  the HERAPDF1.5 PDFs. 
}
\label{fig-xsing}
\vfill
\end{figure}

\begin{figure}[p]
\vfill
\begin{center}           
            \includegraphics[width=1.0\textwidth]{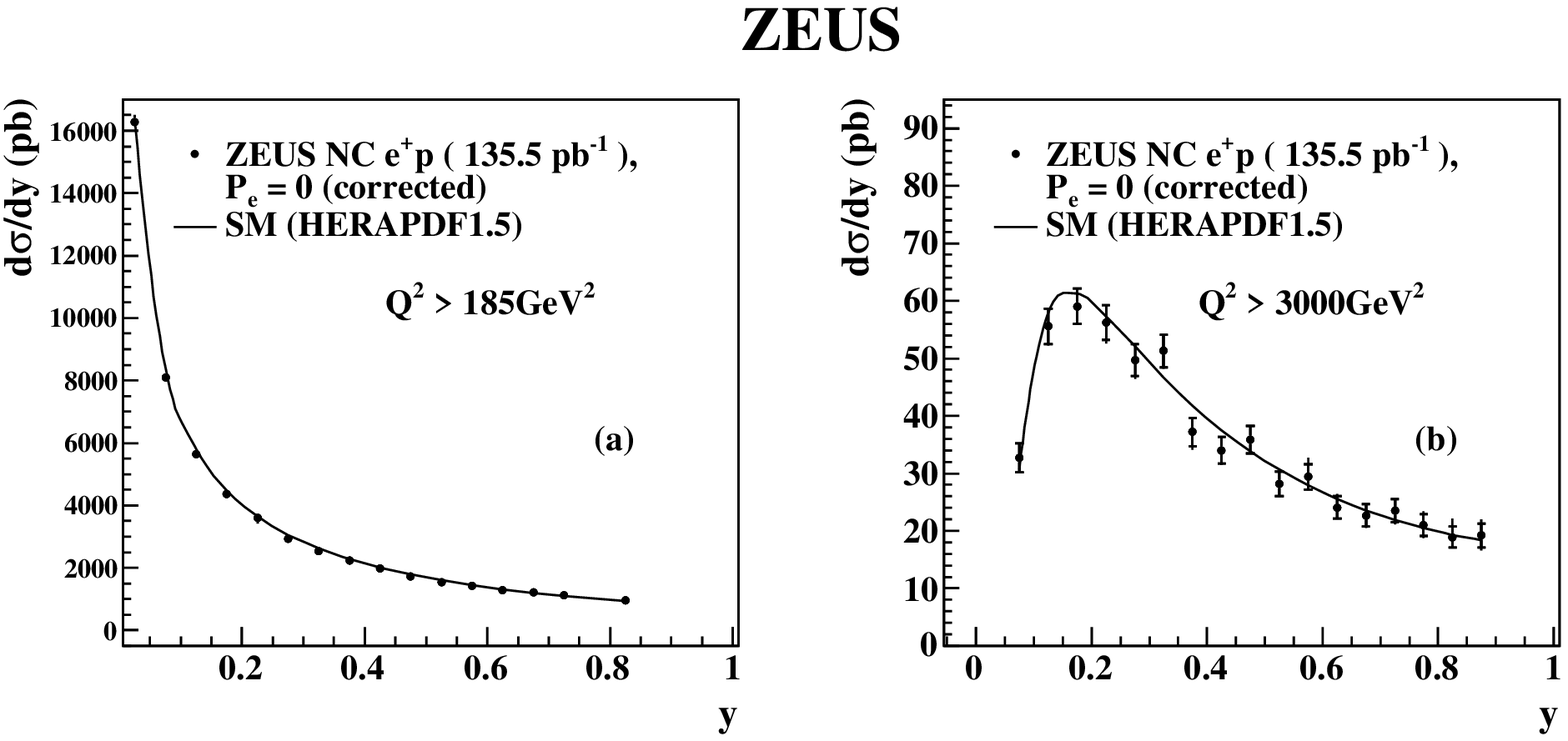}
\end{center}
\caption{
  The $e^+ p$ NC DIS cross-section $d\sigma/dy$ for (a) $Q^2 > 185 \gev^2$ 
  and (b) $Q^2 > 3\;000 \gev^2$ for $y(1-x)^2>0.004$.
  Other details as in Figure~\ref{fig-xsing}. 
}
\label{fig-ysing}
\vfill
\end{figure}

The single-differential cross sections
as a function of $Q^{2}$, $x$ and $y$,
extracted using the full data sample, are 
shown in Figs.~\ref{fig-q2sing}--\ref{fig-ysing} and tabulated in Tables~\ref{tab:dsdq2Total}--\ref{tab:dsdyTotal}. In all tables, the total systematic uncertainty as described in Section~\ref{sec-sys} is given. The numbers for the individual contributions are available electronically~\cite{upub:zeusdat, upub:durham}. 

Combining the data from the negatively and positively polarised beams resulted in a residual polarisation of 0.03 which was corrected for using theoretical predictions in NLO QCD with electroweak corrections. 

The measurement of $d\sigma/dQ^{2}$, shown in \fig{q2sing}, falls over seven orders of magnitude in the measured range covering 
two orders of magnitude in $Q^{2}$. In this figure, the ratio of the measured cross sections and the SM predictions evaluated using the HERAPDF1.5 PDFs~\cite{Abramowicz:2011a, Radescu:2010zz} and the PDFs from ZEUSJETS~\cite{epj:c42:1}, CTEQ6M~\cite{jhep:07:012} and MSTW2008~\cite{Martin:2009iq} are shown. The SM predictions differ depending on the PDFs. Taking into account the luminosity uncertainty, which is not shown in the figures, the data are well described by the SM predictions.
%
The cross-sections $d\sigma/dx$ and
$d\sigma/dy$ 
are shown in \fig{xsing} and \fig{ysing} for the nominal range, $Q^2 > 185 \gev^2$, and for $Q^2 > 3\,000 \gev^2$.
The figures demonstrate the precision of this measurement.
The measured cross sections are well described by the SM
prediction evaluated using the HERAPDF1.5 PDFs.\footnote{HERAPDF1.5 is based on HERA\;I and HERA\;II data, but the data presented here is not used for the extraction.}

\begin{figure}[p]
\vfill
\begin{center}
\includegraphics[width=6in]{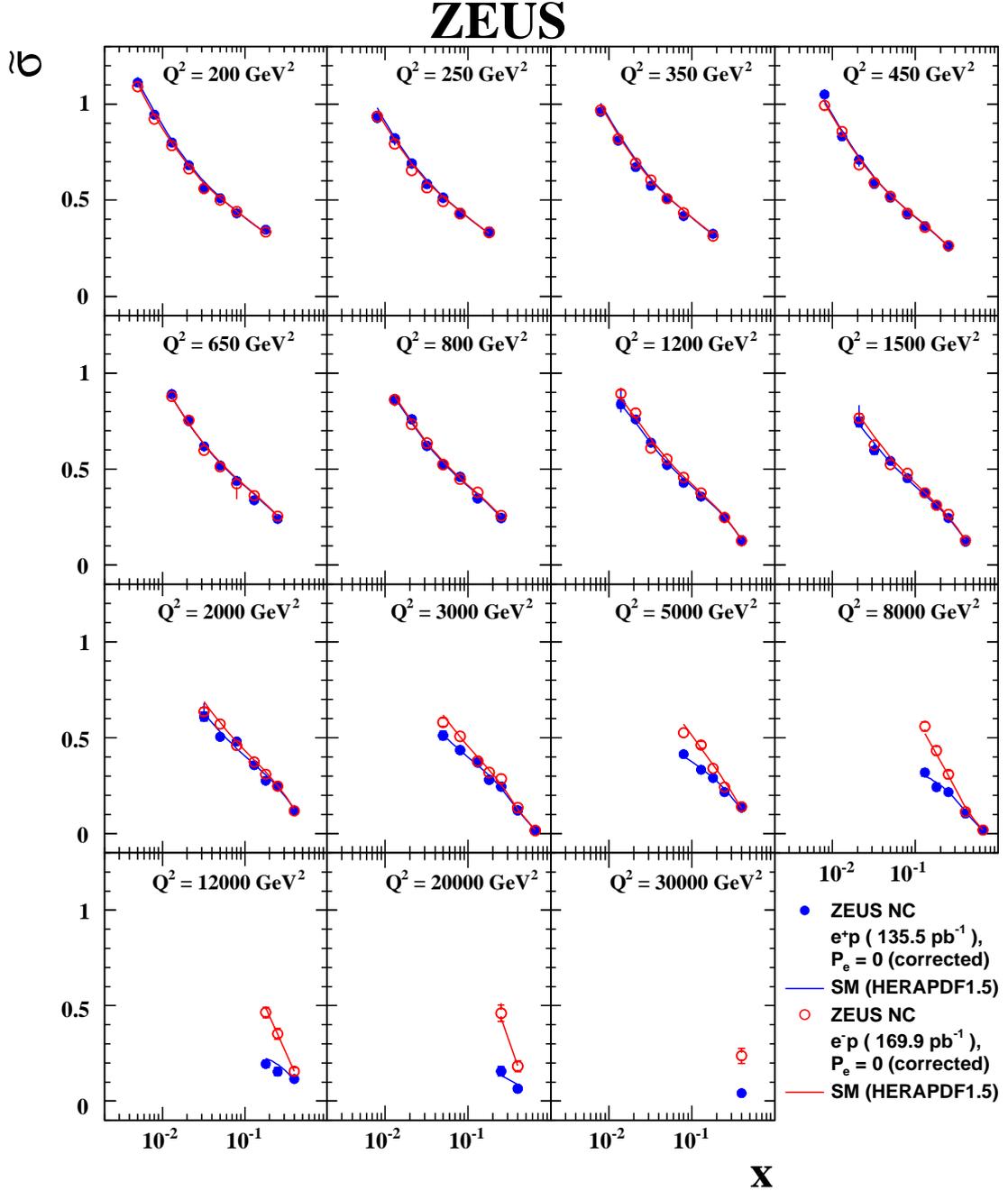}
\end{center}
\caption{
  The $e^\pm p$ unpolarised NC DIS reduced cross-section $\sitil$
  plotted as a function of $x$ at fixed $Q^2$.
  The closed (open) circles represent data points
  for $e^+ p$ ($e^- p$) collisions in which
  the inner error bars show the statistical uncertainty
  while the outer bars show the statistical and systematic uncertainties
  added in quadrature, although errors are too small to be seen in most cases.
  The curves show the predictions of the SM
  evaluated using the HERAPDF1.5 PDFs.
}
\label{fig-red_unpol}
\vfill
\end{figure}

The reduced cross sections of unpolarised $e^ + p$ NC DIS,
tabulated in Table~\ref{tab:ds2dxdq2Total_1},
are shown in \fig{red_unpol}.
The residual polarisation was corrected for using theoretical predictions.
The SM predictions are in good agreement with the measurements
over the full kinematic range. 
Also shown are the unpolarised $e^ - p$ NC DIS cross sections, measured
using an integrated luminosity of $169.9 \pb^{-1}$
collected between 2005 and 2006~\cite{Chekanov:2009gm}.
In Section~\ref{sec-pred}, it was discussed that the $e^-p$ and $e^+p$ reduced cross sections only differ at high $Q^2$. As the contribution of $x\tilde{F_3}$ has to be extracted through a subtraction (see Eq.~(\ref{eqn:xf3})), a very precise measurement of these cross sections is needed.

\begin{figure}[p]
\vfill
\begin{center}
\includegraphics[width=6in]{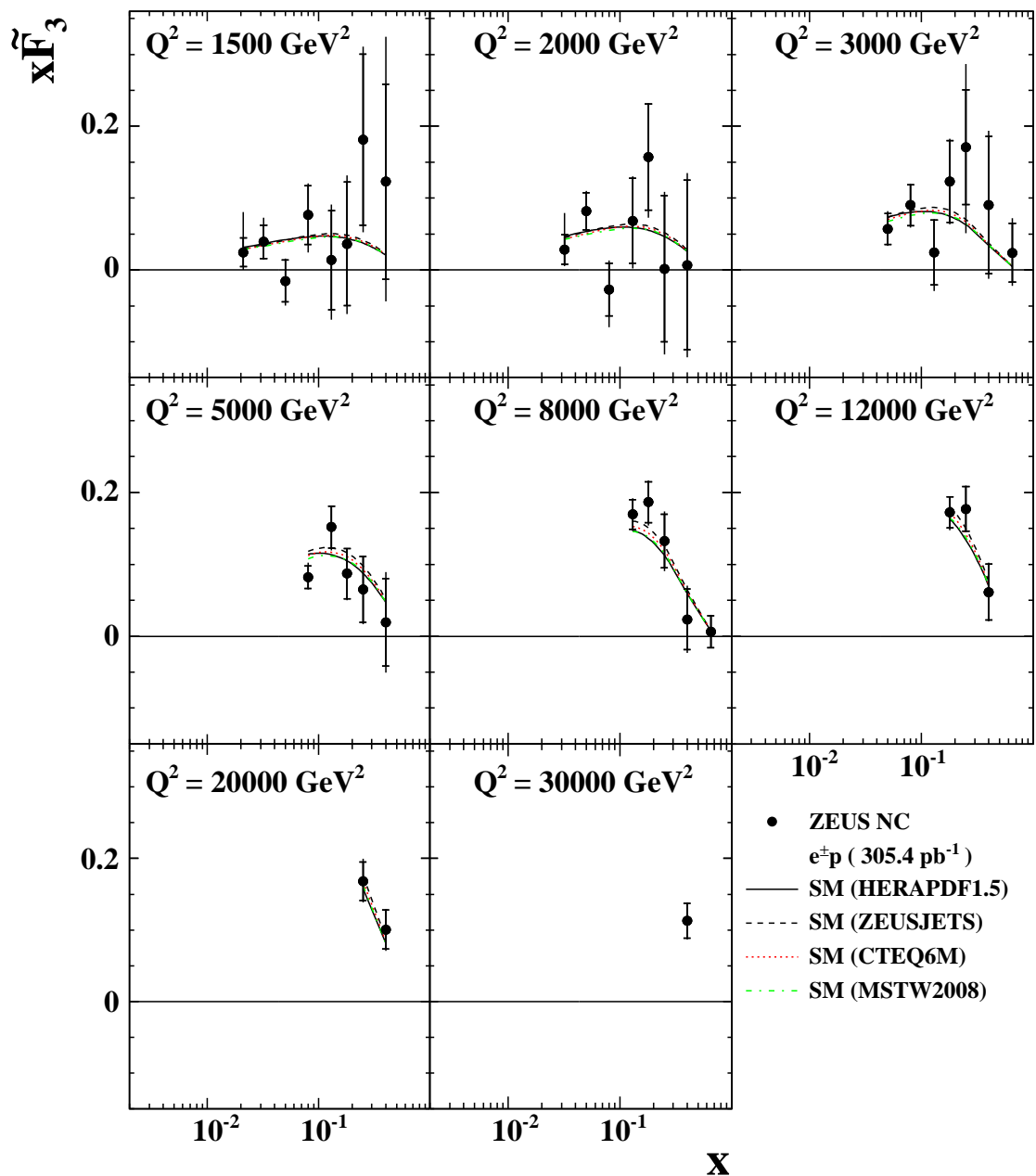}
\end{center}
\caption{
  The structure-function $x\tilde{F_3}$ plotted
  as a function of $x$ at fixed $Q^{2}$.
  The closed circles represent the ZEUS data.
  The inner error bars show the statistical uncertainty
  while the outer ones show the statistical and systematic uncertainties
  added in quadrature.
    The curves show the predictions of the SM
  evaluated using HERAPDF1.5 (solid), ZEUSJETS (dashed), CTEQ6M (dotted) and MSTW2008 (dash-dotted)  PDFs .
}
\label{fig-xf3}
\vfill
\end{figure}

\Fig{xf3} shows the result on $x\tilde{F_3}$
obtained according to Eq.~(\ref{eqn:xf3}) from the unpolarised $e^+ p$ and $e^- p$ reduced cross sections
in the high-$Q^2$ region. The systematic uncertainties were treated as uncorrelated between the $e^+p$ and the $e^-p$ measurements in the extraction of $x\tilde{F_3}$.
The measurements are well described by the SM predictions. The results are also given in Table~\ref{tab:xF3}.

\begin{figure}[p]
\vfill
\begin{center}
\includegraphics[width=6in]{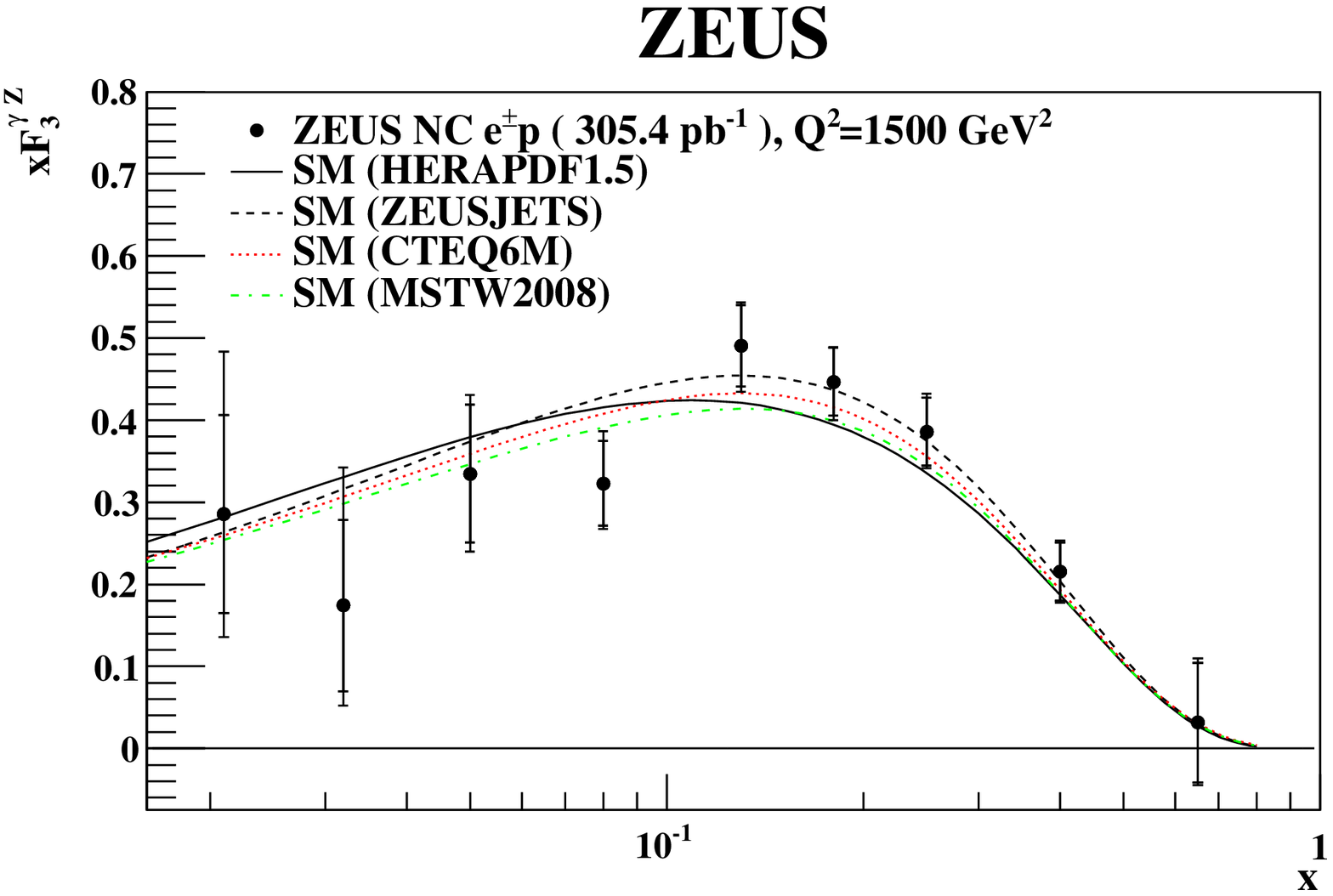} 
\end{center}
\caption{
  The structure function $xF_{3}^{\gamma Z}$ extrapolated
  to a single $Q^2$ value of $1\,500\gev^2$ and plotted as a function of $x$. 
Other details as in Figure~\ref{fig-xf3}.
 }
\label{fig-xf3_gz}
\vfill

\end{figure}

The structure-function $xF_{3}^{\gamma Z}$ has little dependence on $Q^{2}$.
Therefore a higher statistical significance could be obtained by averaging the measurements after an extrapolation to $1\,500 \gev^{2}$. The structure-function $xF_{3}^{\gamma Z}$ measured at $Q^2 = 1\,500 \gev^{2}$,
tabulated in Table~\ref{tab:xF3gz},
is shown in \fig{xf3_gz}.
It is well described by the SM predictions.

The inclusive cross sections presented here provide valuable information
to the global fits~\cite{Nadolsky:2008zw,Martin:2009iq}
for parton distribution functions over a wide range of Bjorken $x$ values from $\sim 10^{-2}$ to $0.65$.

\subsection{Polarised cross sections}
\begin{figure}[p]
\vfill
\begin{center}
\includegraphics[width=6in]{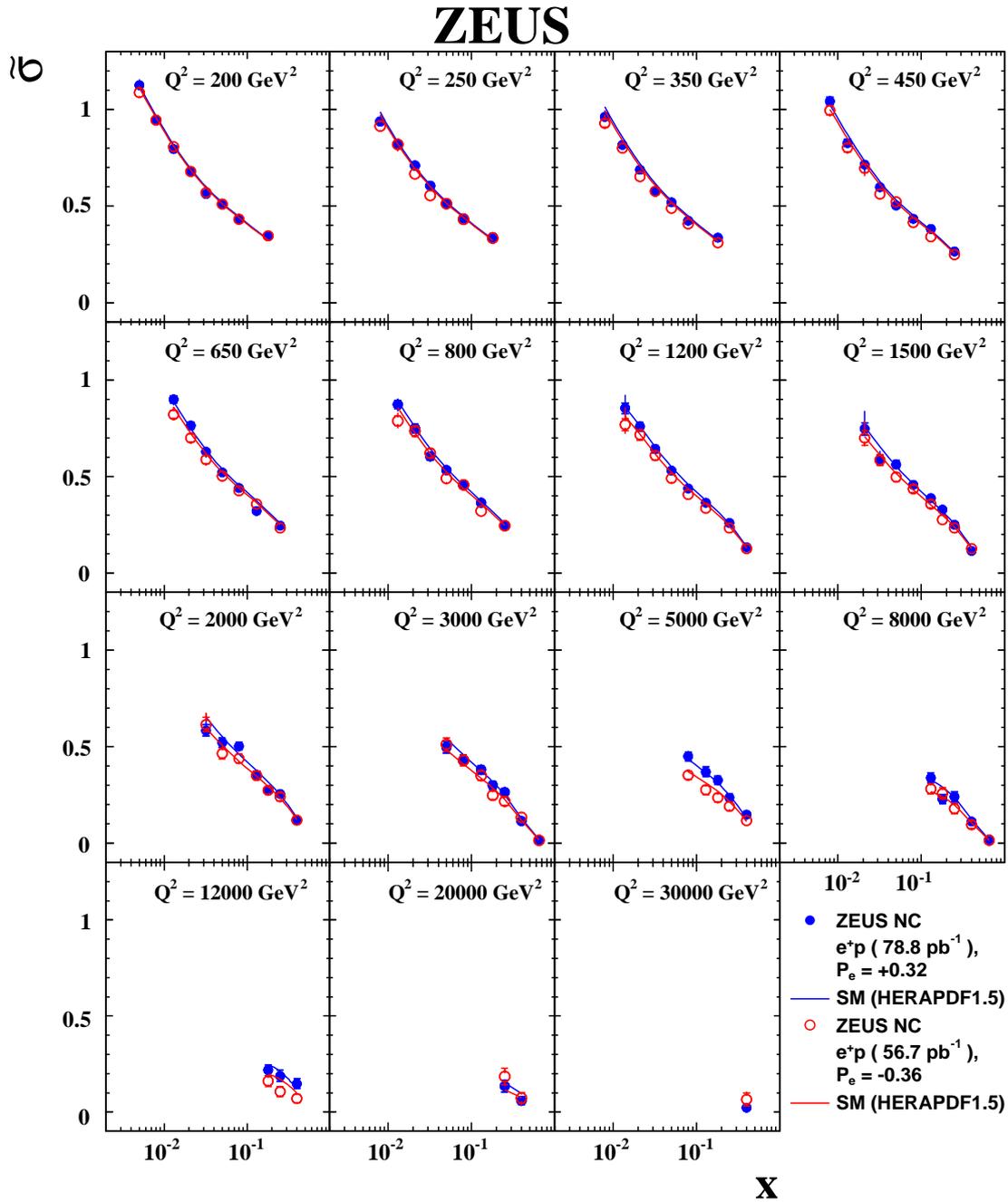}
\end{center}
\caption{
  The $e^+p$ NC DIS reduced cross-section $\sitil$
  for positively and negatively polarised beams
  plotted as a function of $x$ at fixed $Q^2$.
  The closed (open) circles represent the ZEUS data
  for negative (positive) polarisation.
Other details as in Figure~\ref{fig-red_unpol}.
}
\label{fig-red_pol}
\vfill
\end{figure}

\begin{figure}[p]
\vfill
\begin{center}
\includegraphics[width=1.0\textwidth]{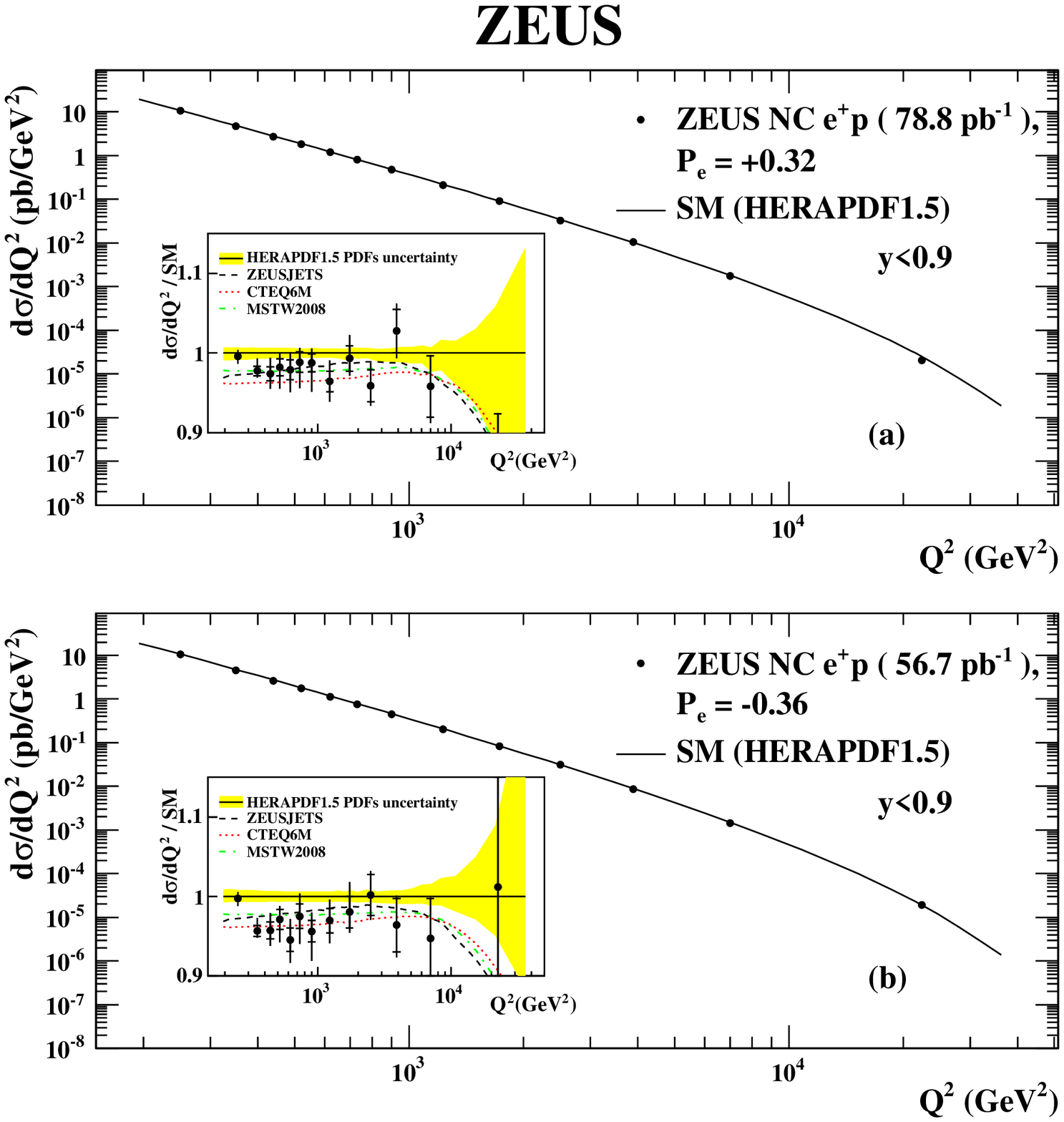}
\end{center}

\caption{
  The $e^+ p$ NC DIS cross-section $d\sigma/dQ^2$ for $y < 0.9$ and $y(1-x)^2>0.004$
  for (a) positive and (b) negative polarisation.
   Other details as in Figure~\ref{fig-q2sing}.
}
\label{fig-dsdq2}
\vfill
\end{figure}

The effects of the longitudinal polarisation of the electrons becomes significant at the electroweak scale,
where the contributions of both $\gamma$ and $Z$ exchange
to the cross section are comparable.
The reduced cross sections
for positive and negative longitudinal polarisations,
tabulated in Tables~\ref{tab:ds2dxdq2Rh_1} to \ref{tab:ds2dxdq2Lh_1},
are shown in \fig{red_pol}. The data are also well described by the SM predictions using the HERAPDF1.5 PDFs.
At high $Q^2$, a difference between the positively and negatively
polarised cross sections is predicted.
To demonstrate this effect,
the single-differential cross-section $d\sigma/dQ^{2}$ 
was measured separately for positive and negative beam polarisations.
The results are  shown in \fig{dsdq2}. 
Both measurements are well described by the SM predictions using different sets of PDFs taking the uncertainty due to the luminosity measurement into account. 
The single-differential cross sections
as a function of $Q^{2}$, $x$ and $y$,
extracted using the negatively and positively polarised data samples separately, are 
tabulated in Tables~\ref{tab:dsdq2Rh} to \ref{tab:dsdyLh}. 

\begin{figure}[p]
\vfill
\begin{center}
\includegraphics[width=1.0\textwidth]{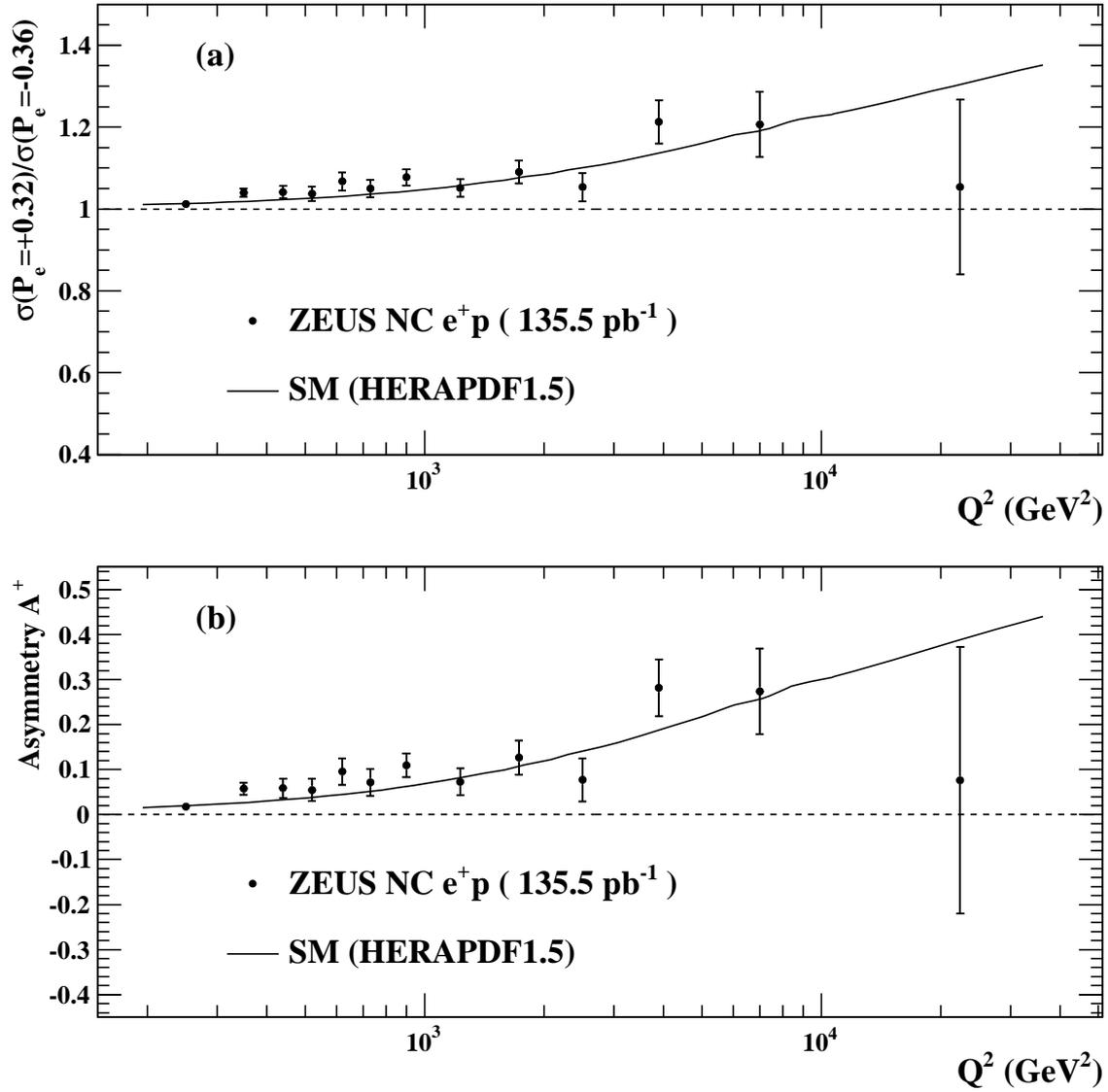}
\end{center}

\caption{
  The (a) ratio of $d\sigma/dQ^{2}$ using positive and negative
  polarisation
  and (b) the polarisation asymmetry $A^{+}$ as functions of $Q^2$.
  The closed circles represent ZEUS data.
  Only statistical uncertainties are considered
  as the systematic uncertainties are assumed to cancel.
  The curves show the predictions of the SM
  evaluated using the HERAPDF1.5 PDFs.
}
\label{fig-asym}
\vfill
\end{figure}

The ratio of the measured single-differential cross-section $d\sigma/dQ^{2}$ 
for the two different polarisation states
is shown in \fig{asym} (a).
The difference between the two polarisation states
is clearly visible at higher $Q^{2}$.
The asymmetry $A^{+}$ (see Eq.~(\ref{eqn:asymMeas}))
extracted from these measurements
is tabulated in Table~\ref{tab:asym}
and is shown in \fig{asym} (b),
where only statistical uncertainties are considered.
The uncertainty in $A^{+}$ arising from the relative normalization
between the data sets was evaluated to be $1\,\%$.
The other systematic uncertainties are assumed to cancel.
The SM also describes these results well. 
The deviation of $A^{+}$ from zero,
particularly at high $Q^2$,
shows the difference in the behaviour of the two polarisation states
and is clear evidence of parity violation. The precision of the data makes the effect also clearly visible at relatively low $Q^2$, where it is intrinsically small.

The effect of $\gamma$/$Z$ interference is quantified
by calculating the $\chi^2$ per degree of freedom of $A^{+}$
with respect both to zero and to the SM prediction
using the HERAPDF1.5 PDFs.
The $\chi^2/\rm{d.o.f.}$ with respect to zero is determined to be 9.0,
whereas the $\chi^2/\rm{d.o.f.}$ with respect to the SM prediction is 1.5.
Thus parity violation in $e p$ NC DIS is
demonstrated at scales down to $\approx 10^{-18}$ m. 

The polarised cross sections presented here constrain the vector couplings 
of the quarks to the $Z$ (see Eq.~(\ref{fgf})) when included in the PDF fits.
Therefore, this measurement is
a stringent test of the electroweak sector of the Standard Model.
The data can also be used to test physics beyond the Standard Model 
like setting limits on the production of leptoquarks~\cite{Abramovic:2012}.

\section{Conclusions}
\label{sec-conclusion}

The cross sections for neutral current deep inelastic $e^+p$ scattering
 with a longitudinally polarised positron beam have been measured.
The measurements are based on a data sample corresponding to an
 integrated luminosity of
$135.5 \pb^{-1}$ collected with the ZEUS detector at HERA from 2006 to 2007
at a centre-of-mass energy of $318 \gev$.
The accessible range in $Q^2$ extended to $Q^2 = 50\,000 \gev^2$, allowing for a stringent test of electroweak effects in the Standard Model.

The single-differential cross sections as a function of $Q^2$, $x$ and $y$
were presented for $Q^2 > 185 \gev^2$, $y < 0.9$ and $y(1-x)^2>0.004$,
where the data obtained with negatively and positively polarised beams
were combined.
The cross-sections $d\sigma/dx$ and $d\sigma/dy$
were also measured for $Q^2 > 3\,000 \gev^2$, $y < 0.9$ and $y(1-x)^2>0.004$.  
The reduced cross section was measured 
at zero polarisation
by correcting for the residual polarisation of the combined data sample.
These measurements were combined with previously measured
$e^-p$ neutral current cross sections to extract $x\tilde{F_3}$.
In addition, the interference structure function $xF_{3}^{\gamma Z}$
was extracted at an average value of $Q^{2} = 1\,500 \gev^2$.

The reduced cross section and
the single-differential cross-sections $d\sigma/dQ^2$, $d\sigma/dx$ and $d\sigma/dy$
 were also measured separately for positive and negative values
of the longitudinal polarisation of the positron beam.
Parity violation was observed through the polarisation asymmetry $A^+$.
The measured cross sections confirm the predictions of the Standard Model
and provide strong constraints at the electroweak scale.

\section*{Acknowledgements}
\label{sec-ack}

\Zacknowledge

\vfill\eject

{
\def\bibname{\Large\bf References}
\def\refname{\Large\bf References}
\pagestyle{plain}
\ifzeusbst
  \bibliographystyle{l4z_default}
\fi
\ifzdrftbst
  \bibliographystyle{l4z_draft}
\fi
\ifzbstepj
  \bibliographystyle{l4z_epj}
\fi
\ifzbstnp
  \bibliographystyle{l4z_np}
\fi
\ifzbstpl
  \bibliographystyle{l4z_pl}
\fi
{\raggedright
\bibliography{syn.bib,%
              l4z_articles.bib,%
              l4z_books.bib,%
              l4z_conferences.bib,%
              l4z_h1.bib,%
              l4z_misc.bib,%
              l4z_old.bib,%
              l4z_preprints.bib,%
              l4z_replaced.bib,%
              l4z_temporary.bib,%
              l4z_zeus.bib}}
}
\vfill\eject
\newpage
\begin{table}
\begin{scriptsize}
\begin{center} \begin{tabular}[t]{|c|r|l|r|r|} \hline
\multicolumn{1}{|c|}{$Q^2$ range} & \multicolumn{1}{c|}{$Q^2_c$} & \multicolumn{1}{c|}{$d\sigma / dQ^{2}$} & \multicolumn{1}{c|}{$N_{\text{data}}$} & \multicolumn{1}{c|}{$N^{\text{MC}}_{\text{bg}}$}\\
\multicolumn{1}{|c|}{($\gev^{2}$)} & \multicolumn{1}{c|}{($\gev^{2}$)} & \multicolumn{1}{c|}{($\pb / \gev^{2}$)} & \multicolumn{1}{c|}{} & \multicolumn{1}{c|}{}\\ \hline \hline
$185 - 210$ & 195 & ($1.91 \pm 0.01 ^{+0.02}_{-0.01}$) $\times$ $10^{1}$ & 55281 & 110.1 \\ 
$210 - 240$ & 220 & ($1.43 \pm 0.01 ^{+0.01}_{-0.01}$) $\times$ $10^{1}$ & 47861 & 73.8 \\ 
$240 - 270$ & 255 & ($1.01 \pm 0.01 ^{+0.01}_{-0.01}$) $\times$ $10^{1}$ & 34808 & 59.2 \\ 
$270 - 300$ & 285 & ($7.79 \pm 0.05 ^{+0.07}_{-0.09}$)  & 25835 & 18.9 \\ 
$300 - 340$ & 320 & ($5.79 \pm 0.04 ^{+0.09}_{-0.05}$)  & 24184 & 32.8 \\ 
$340 - 380$ & 360 & ($4.35 \pm 0.03 ^{+0.07}_{-0.02}$)  & 17201 & 22.8 \\ 
$380 - 430$ & 400 & ($3.33 \pm 0.03 ^{+0.06}_{-0.03}$)  & 15791 & 28.5 \\ 
$430 - 480$ & 450 & ($2.56 \pm 0.02 ^{+0.05}_{-0.05}$)  & 11903 & 40.1 \\ 
$480 - 540$ & 510 & ($1.89 \pm 0.02 ^{+0.02}_{-0.05}$)  & 10365 & 19.9 \\ 
$540 - 600$ & 570 & ($1.39 \pm 0.02 ^{+0.03}_{-0.03}$)  & 6943 & 36.2 \\ 
$600 - 670$ & 630 & ($1.14 \pm 0.01 ^{+0.02}_{-0.03}$)  & 6366 & 20.1 \\ 
$670 - 740$ & 700 & ($8.70 \pm 0.12 ^{+0.19}_{-0.29}$) $\times$ $10^{-1}$ & 5655 & 22.5 \\ 
$740 - 820$ & 780 & ($6.65 \pm 0.09 ^{+0.09}_{-0.19}$) $\times$ $10^{-1}$ & 5750 & 16.9 \\ 
$820 - 900$ & 860 & ($5.08 \pm 0.07 ^{+0.09}_{-0.17}$) $\times$ $10^{-1}$ & 4654 & 25.0 \\ 
$900 - 990$ & 940 & ($4.16 \pm 0.06 ^{+0.06}_{-0.14}$) $\times$ $10^{-1}$ & 4295 & 15.6 \\ 
$990 - 1080$ & 1030 & ($3.20 \pm 0.06 ^{+0.09}_{-0.13}$) $\times$ $10^{-1}$ & 3304 & 10.5 \\ 
$1080 - 1200$ & 1130 & ($2.55 \pm 0.04 ^{+0.04}_{-0.06}$) $\times$ $10^{-1}$ & 3522 & 18.1 \\ 
$1200 - 1350$ & 1270 & ($1.96 \pm 0.03 ^{+0.06}_{-0.05}$) $\times$ $10^{-1}$ & 3439 & 14.6 \\ 
$1350 - 1500$ & 1420 & ($1.42 \pm 0.03 ^{+0.03}_{-0.03}$) $\times$ $10^{-1}$ & 2501 & 16.4 \\ 
$1500 - 1700$ & 1590 & ($1.08 \pm 0.02 ^{+0.03}_{-0.02}$) $\times$ $10^{-1}$ & 2549 & 8.3 \\ 
$1700 - 1900$ & 1790 & ($7.84 \pm 0.18 ^{+0.21}_{-0.13}$) $\times$ $10^{-2}$ & 1849 & 8.5 \\ 
$1900 - 2100$ & 1990 & ($5.88 \pm 0.16 ^{+0.21}_{-0.10}$) $\times$ $10^{-2}$ & 1393 & 9.1 \\ 
$2100 - 2600$ & 2300 & ($4.02 \pm 0.08 ^{+0.08}_{-0.08}$) $\times$ $10^{-2}$ & 2311 & 7.2 \\ 
$2600 - 3200$ & 2800 & ($2.34 \pm 0.06 ^{+0.03}_{-0.04}$) $\times$ $10^{-2}$ & 1565 & 3.3 \\ 
$3200 - 3900$ & 3500 & ($1.31 \pm 0.04 ^{+0.03}_{-0.03}$) $\times$ $10^{-2}$ & 1083 & 1.1 \\ 
$3900 - 4700$ & 4200 & ($7.77 \pm 0.29 ^{+0.14}_{-0.14}$) $\times$ $10^{-3}$ & 715 & 3.9 \\ 
$4700 - 5600$ & 5100 & ($4.18 \pm 0.20 ^{+0.04}_{-0.12}$) $\times$ $10^{-3}$ & 447 & 0.0 \\ 
$5600 - 6600$ & 6050 & ($2.66 \pm 0.15 ^{+0.03}_{-0.06}$) $\times$ $10^{-3}$ & 320 & 0.0 \\ 
$6600 - 7800$ & 7100 & ($1.47 \pm 0.10 ^{+0.04}_{-0.06}$) $\times$ $10^{-3}$ & 208 & 0.0 \\ 
$7800 - 9200$ & 8400 & ($9.20 \pm 0.74 ^{+0.31}_{-0.34}$) $\times$ $10^{-4}$ & 152 & 0.0 \\ 
$9200 - 12800$ & 10800 & ($3.40 \pm 0.28 ^{+0.09}_{-0.10}$) $\times$ $10^{-4}$ & 145 & 0.0 \\ 
$12800 - 18100$ & 15200 & ($9.21 \pm 1.21 ^{+0.31}_{-0.76}$) $\times$ $10^{-5}$ & 57 & 0.0 \\ 
$18100 - 25600$ & 21500 & ($3.81 ^{+0.76}_{-0.64}$ $^{+0.23}_{-0.23}$) $\times$ $10^{-5}$ & 35 & 0.0 \\ 
$25600 - 50000$ & 36200 & ($8.23 ^{+6.51}_{-3.94}$ $^{+0.71}_{-0.44}$) $\times$ $10^{-7}$ & 4 & 0.0 \\ 
\hline
\end{tabular}
\end{center}
\caption[]
{The single-differential cross-section $d\sigma / dQ^{2}$ ($y < 0.9$, $y(1-x)^2>0.004$)
for the reaction $e^{+}p \rightarrow e^{+}X$ ($\mathcal{L} = 135.5 \pbi$, corrected to $P_{e} = 0$).
The bin range, bin centre ($Q^2_c$) and measured cross section
corrected to the electroweak Born level are shown.
The first (second) error on the cross section
corresponds to the statistical (systematic) uncertainties.
The number of observed data events ($N_{\text{data}}$)
and simulated background events ($N^{\text{MC}}_{\text{bg}}$) are also shown.}
\label{tab:dsdq2Total}
\end{scriptsize}
\end{table}
\newpage
\begin{table}
\begin{scriptsize}
\begin{center} \begin{tabular}[t]{|r|l|l|l|r|r|} \hline
\multicolumn{1}{|c|}{$Q^2~>$} & \multicolumn{1}{c|}{$x$ range} & \multicolumn{1}{c|}{$x_c$} & \multicolumn{1}{c|}{$d\sigma / dx$} & \multicolumn{1}{c|}{$N_{\text{data}}$} & \multicolumn{1}{c|}{$N^{\text{MC}}_{\text{bg}}$}\\
\multicolumn{1}{|c|}{($\gev^2$)} & \multicolumn{1}{c|}{} & \multicolumn{1}{c|}{} & \multicolumn{1}{c|}{($\pb$)} & & \\
\hline \hline
185 & $(0.63 - 1.00)\times 10^{-2}$ & $0.794\times 10^{-2}$ & ($8.71 \pm 0.05 ^{+0.13}_{-0.14}$) $\times$ $10^{4}$ & 34570 & 161.0 \\ 
 & $(0.10 - 0.16)\times 10^{-1}$ & $0.126\times 10^{-1}$ & ($5.84 \pm 0.03 ^{+0.07}_{-0.17}$) $\times$ $10^{4}$ & 39862 & 122.5 \\ 
 & $(0.16 - 0.25)\times 10^{-1}$ & $0.200\times 10^{-1}$ & ($3.63 \pm 0.02 ^{+0.03}_{-0.03}$) $\times$ $10^{4}$ & 39233 & 82.9 \\ 
 & $(0.25 - 0.40)\times 10^{-1}$ & $0.316\times 10^{-1}$ & ($2.10 \pm 0.01 ^{+0.02}_{-0.01}$) $\times$ $10^{4}$ & 38384 & 30.2 \\ 
 & $(0.40 - 0.63)\times 10^{-1}$ & $0.501\times 10^{-1}$ & ($1.24 \pm 0.01 ^{+0.01}_{-0.01}$) $\times$ $10^{4}$ & 33557 & 5.5 \\ 
 & $(0.63 - 1.00)\times 10^{-1}$ & $0.794\times 10^{-1}$ & ($6.90 \pm 0.04 ^{+0.08}_{-0.03}$) $\times$ $10^{3}$ & 31825 & 5.1 \\ 
 & $0.10 - 0.16$ & $0.126$ & ($3.89 \pm 0.02 ^{+0.04}_{-0.02}$) $\times$ $10^{3}$ & 30244 & 1.4 \\ 
 & $0.16 - 0.25$ & $0.200$ & ($2.04 \pm 0.01 ^{+0.04}_{-0.06}$) $\times$ $10^{3}$ & 18768 & 0.0 \\ 
\hline
3000 & $(0.40 - 0.63)\times 10^{-1}$ & $0.501\times 10^{-1}$ & ($1.71 \pm 0.08 ^{+0.06}_{-0.03}$) $\times$ $10^{2}$ & 440 & 1.1 \\ 
 & $(0.63 - 1.00)\times 10^{-1}$ & $0.794\times 10^{-1}$ & ($1.60 \pm 0.06 ^{+0.04}_{-0.02}$) $\times$ $10^{2}$ & 714 & 3.9 \\ 
 & $0.10 - 0.16$ & $0.126$ & ($1.18 \pm 0.04 ^{+0.01}_{-0.04}$) $\times$ $10^{2}$ & 859 & 0.0 \\ 
 & $0.16 - 0.25$ & $0.200$ & ($6.72 \pm 0.25 ^{+0.06}_{-0.16}$) $\times$ $10^{1}$ & 730 & 0.0 \\ 
 & $0.25 - 0.40$ & $0.316$ & ($3.22 \pm 0.14 ^{+0.04}_{-0.08}$) $\times$ $10^{1}$ & 567 & 0.0 \\ 
 & $0.40 - 0.75$ & $0.687$ & ($1.20 \pm 0.08 ^{+0.02}_{-0.02}$)  & 240 & 0.0 \\ 
\hline
\end{tabular}
\end{center}
\caption[]
{The single-differential cross-section $d\sigma / dx$ ($y<0.9$, $y(1-x)^2>0.004$)
for $Q^2 > 185 \gev^2$ and $Q^2 > 3\,000 \gev^2$
for the reaction $e^{+}p \rightarrow e^{+}X$ ($\mathcal{L} = 135.5 \pbi,$ corrected to $P_{e} = 0$).
The $Q^2$ and bin range, bin centre ($x_c$) and measured cross section
corrected to the electroweak Born level are shown. Other details as in Table~\ref{tab:dsdq2Total}.}
\label{tab:dsdxTotal}
\end{scriptsize}
\end{table}
\newpage
\begin{table}
\begin{scriptsize}
\begin{center} 
\begin{tabular}[t]{|r|r|r|r|r|r|} \hline
\multicolumn{1}{|c|}{$Q^{2}~>$} & \multicolumn{1}{c|}{$y$ range} & \multicolumn{1}{c|}{$y_c$} & \multicolumn{1}{c|}{$d\sigma / dy$} & \multicolumn{1}{c|}{$N_{\text{data}}$} & \multicolumn{1}{c|}{$N^{\text{MC}}_{\text{bg}}$}\\
\multicolumn{1}{|c|}{($\gev^{2}$)} & \multicolumn{1}{c|}{} & \multicolumn{1}{c|}{} & \multicolumn{1}{c|}{($\pb$)} & & \\
\hline \hline
185 & $0.00 - 0.05$ & 0.025 & ($1.63 \pm 0.01 ^{+0.02}_{-0.01}$) $\times$ $10^{4}$ & 75314 & 0.0 \\ 
 & $0.05 - 0.10$ & 0.075 & ($8.10 \pm 0.04 ^{+0.09}_{-0.05}$) $\times$ $10^{3}$ & 50532 & 7.4 \\ 
 & $0.10 - 0.15$ & 0.125 & ($5.64 \pm 0.03 ^{+0.03}_{-0.05}$) $\times$ $10^{3}$ & 34944 & 11.6 \\ 
 & $0.15 - 0.20$ & 0.175 & ($4.37 \pm 0.03 ^{+0.02}_{-0.14}$) $\times$ $10^{3}$ & 26237 & 24.7 \\ 
 & $0.20 - 0.25$ & 0.225 & ($3.61 \pm 0.03 ^{+0.03}_{-0.20}$) $\times$ $10^{3}$ & 20974 & 19.8 \\ 
 & $0.25 - 0.30$ & 0.275 & ($2.93 \pm 0.02 ^{+0.04}_{-0.04}$) $\times$ $10^{3}$ & 16254 & 17.7 \\
\hline
\end{tabular}
\end{center}
\caption[]
{The single-differential cross-section $d\sigma / dy$
for $Q^2 > 185 \gev^2$ and $Q^2 > 3\,000 \gev^2$ ($y(1-x)^2>0.004$)
for the reaction $e^{+}p \rightarrow e^{-}X$ ($\mathcal{L} = 135.5\pbi$, corrected to $P_{e} = 0$).
The $Q^2$ and bin range, bin centre ($y_c$) and measured cross section
corrected to the electroweak Born level are shown.
Other details as in Table~\ref{tab:dsdq2Total}.}
\label{tab:dsdyTotal}
\end{scriptsize}
\end{table}
\newpage
\begin{table} \begin{scriptsize} \begin{center} \begin{tabular}[t]{|l|r|l|l|l|r|r|} \hline
\multicolumn{1}{|c|}{$Q^2$ range} & \multicolumn{1}{c|}{$Q^2_c$} & \multicolumn{1}{|c|}{$x$ range} & \multicolumn{1}{c|}{$x_c$} & \multicolumn{1}{c|}{$\tilde{\sigma}$} & \multicolumn{1}{c|}{$N_{\text{data}}$} & \multicolumn{1}{c|}{$N^{\text{MC}}_{\text{bg}}$} \\
\multicolumn{1}{|c|}{($\gev^{2}$)} & \multicolumn{1}{c|}{($\gev^{2}$)} & \multicolumn{1}{|c|}{} & \multicolumn{1}{c|}{} & \multicolumn{1}{c|}{} & & \\ \hline \hline
$185 - 240$ & 200 & $0.004 - 0.006$ & 0.005 & ($1.110 \pm 0.010 ^{+0.029}_{-0.017}$) & 13313 & 108.0 \\ 
 & 200 & $0.006 - 0.010$ & 0.008 & ($0.945 \pm 0.008 ^{+0.008}_{-0.022}$) & 15647 & 38.8 \\ 
 & 200 & $0.010 - 0.017$ & 0.013 & ($0.801 \pm 0.006 ^{+0.002}_{-0.022}$) & 16074 & 14.1 \\ 
 & 200 & $0.017 - 0.025$ & 0.021 & ($0.680 \pm 0.006 ^{+0.013}_{-0.004}$) & 11107 & 5.1 \\ 
 & 200 & $0.025 - 0.037$ & 0.032 & ($0.566 \pm 0.006 ^{+0.015}_{-0.001}$) & 9767 & 0.0 \\ 
 & 200 & $0.037 - 0.060$ & 0.050 & ($0.511 \pm 0.005 ^{+0.011}_{-0.002}$) & 10375 & 1.1 \\
\hline
\end{tabular}
\end{center}
\end{scriptsize}
\caption[]
{The reduced cross-section $\tilde{\sigma}$ for the reaction
$e^{+}p \rightarrow e^{+}X$ ($\mathcal{L} = 135.5 \pbi$, corrected to $P_{e} = 0$).
The bin range, bin centre ($Q^2_c$ and $x_c$)
and measured cross section corrected to the electroweak Born level are shown.
Other details as in Table~\ref{tab:dsdq2Total}.
This table has two continuations.}
\label{tab:ds2dxdq2Total_1}
\end{table}
\begin{table} \begin{scriptsize} \begin{center} \begin{tabular}[t]{|l|r|l|l|l|r|r|} \hline
\multicolumn{1}{|c|}{$Q^2$ range} & \multicolumn{1}{c|}{$Q^2_c$} & \multicolumn{1}{|c|}{$x$ range} & \multicolumn{1}{c|}{$x_c$} & \multicolumn{1}{c|}{$\tilde{\sigma}$} & \multicolumn{1}{c|}{$N_{\text{data}}$} & \multicolumn{1}{c|}{$N^{\text{MC}}_{\text{bg}}$} \\
\multicolumn{1}{|c|}{($\gev^{2}$)} & \multicolumn{1}{c|}{($\gev^{2}$)} & \multicolumn{1}{|c|}{} & \multicolumn{1}{c|}{} & \multicolumn{1}{c|}{} & & \\ \hline \hline
$530 - 710$ & 650 & $0.010 - 0.017$ & 0.013 & ($0.865 \pm 0.014 ^{+0.024}_{-0.025}$) & 4045 & 29.0 \\ 
 & 650 & $0.017 - 0.025$ & 0.021 & ($0.735 \pm 0.015 ^{+0.008}_{-0.018}$) & 2564 & 0.0 \\ 
 & 650 & $0.025 - 0.037$ & 0.032 & ($0.609 \pm 0.013 ^{+0.009}_{-0.025}$) & 2043 & 1.1 \\ 
 & 650 & $0.037 - 0.060$ & 0.050 & ($0.512 \pm 0.011 ^{+0.003}_{-0.010}$) & 2028 & 0.0 \\ 
 & 650 & $0.060 - 0.100$ & 0.080 & ($0.434 \pm 0.010 ^{+0.002}_{-0.011}$) & 1809 & 0.0 \\ 
 & 650 & $0.100 - 0.170$ & 0.130 & ($0.335 \pm 0.008 ^{+0.005}_{-0.003}$) & 1598 & 0.0 \\
\hline
\end{tabular}
\end{center}
\end{scriptsize}
\vspace{0.5cm} \hspace{5.4cm} {\bf Table 4:}~~~{\it Continuation 1.}
\label{tab:ds2dxdq2Total_2}
\end{table}
\begin{table} \begin{scriptsize} \begin{center} \begin{tabular}[t]{|l|r|l|l|l|r|r|} \hline
\multicolumn{1}{|c|}{$Q^2$ range} & \multicolumn{1}{c|}{$Q^2_c$} & \multicolumn{1}{|c|}{$x$ range} & \multicolumn{1}{c|}{$x_c$} & \multicolumn{1}{c|}{$\tilde{\sigma}$} & \multicolumn{1}{c|}{$N_{\text{data}}$} & \multicolumn{1}{c|}{$N^{\text{MC}}_{\text{bg}}$} \\
\multicolumn{1}{|c|}{($\gev^{2}$)} & \multicolumn{1}{c|}{($\gev^{2}$)} & \multicolumn{1}{|c|}{} & \multicolumn{1}{c|}{} & \multicolumn{1}{c|}{} & & \\ \hline \hline
$2500 - 3500$ & 3000 & $0.037 - 0.060$ & 0.050 & ($0.500 \pm 0.022 ^{+0.012}_{-0.012}$) & 502 & 2.2 \\ 
 & 3000 & $0.060 - 0.100$ & 0.080 & ($0.429 \pm 0.018 ^{+0.004}_{-0.010}$) & 575 & 0.0 \\ 
 & 3000 & $0.100 - 0.150$ & 0.130 & ($0.366 \pm 0.017 ^{+0.003}_{-0.014}$) & 448 & 0.0 \\ 
 & 3000 & $0.150 - 0.230$ & 0.180 & ($0.276 \pm 0.015 ^{+0.004}_{-0.005}$) & 356 & 0.0 \\ 
 & 3000 & $0.230 - 0.350$ & 0.250 & ($0.243 \pm 0.014 ^{+0.002}_{-0.008}$) & 286 & 0.0 \\ 
 & 3000 & $0.350 - 0.530$ & 0.400 & ($0.121 \pm 0.011 ^{+0.004}_{-0.003}$) & 127 & 0.0 \\
\hline
\end{tabular}
\end{center}
\end{scriptsize}
\vspace{0.5cm} \hspace{5.4cm} {\bf Table 4:}~~~{\it Continuation 2.}
\label{tab:ds2dxdq2Total_3}
\end{table}
\clearpage
\newpage
\begin{table} 
\begin{footnotesize}
\begin{center}
\begin{tabular}[t]{|c|r|c|l|c|} \hline
\multicolumn{1}{|c|}{$Q^2$ range} & \multicolumn{1}{c|}{$Q^{2}_c$} & \multicolumn{1}{c|}{$x$ range} & \multicolumn{1}{c|}{$x_c$} & \multicolumn{1}{c|}{$x\tilde{F_3} \times 10$} \\
\multicolumn{1}{|c|}{($\gev^{2}$)} & \multicolumn{1}{c|}{($\gev^{2}$)} & \multicolumn{1}{c|}{} & \multicolumn{1}{c|}{} & \multicolumn{1}{c|}{} \\ \hline \hline
$1300-1800 $ & 1500 & $0.017-0.025 $ & 0.021 & $0.24 \pm 0.20^{+0.52}_{-0.14}$ \\ 
$ $ & 1500 & $0.025-0.037 $ & 0.032 & $0.39 \pm 0.23^{+0.24}_{-0.08}$ \\ 
$ $ & 1500 & $0.037-0.06 $ & 0.050 & $-0.15 \pm 0.29^{+0.11}_{-0.18}$ \\ 
$ $ & 1500 & $0.06-0.1 $ & 0.080 & $0.76 \pm 0.41^{+0.17}_{-0.33}$ \\ 
$ $ & 1500 & $0.1-0.15 $ & 0.130 & $0.14 \pm 0.69^{+0.35}_{-0.46}$ \\ 
$ $ & 1500 & $0.15-0.23 $ & 0.180 & $0.36 \pm 0.86^{+0.40}_{-0.46}$ \\
\hline\end{tabular}
\end{center}
\end{footnotesize}
\caption[]
{The structure-function $x\tilde{F_3}$
extracted using the $e^{+}p$ data set
($\mathcal{L} = 135.5 \pbi$, corrected to $P_{e}=0$)
and previously published NC $e^{-}p$ DIS results
($\mathcal{L} = 169.9 \pbi$, corrected to $P_{e} = 0$).
The bin range and bin centre for $Q^2$ and $x$,
and measured $x\tilde{F_3}$ are shown.
Other details as in Table~\ref{tab:dsdq2Total}.}
\label{tab:xF3}
\end{table}
\begin{table} 
\begin{center}
\begin{tabular}[t]{|r|r|r|}
\hline
\multicolumn{1}{|c|}{$Q^{2}$} & \multicolumn{1}{c|}{$x_c$} & \multicolumn{1}{c|}{$xF_{3}^{\gamma Z} \times 10$} \\
\multicolumn{1}{|c|}{($\gev^{2}$)} & &\\
\hline \hline
1500 & 0.021 & $2.85 \pm 1.21^{+1.57}_{-0.88}$ \\ 
 & 0.032 & $1.74 \pm 1.05^{+1.32}_{-0.62}$ \\ 
 & 0.050 & $3.35 \pm 0.84^{+0.46}_{-0.43}$ \\ 
 & 0.080 & $3.23 \pm 0.52^{+0.37}_{-0.21}$ \\ 
 & 0.130 & $4.91 \pm 0.50^{+0.17}_{-0.26}$ \\ 
 & 0.180 & $4.47 \pm 0.41^{+0.11}_{-0.22}$ \\ 
 & 0.250 & $3.86 \pm 0.41^{+0.20}_{-0.16}$ \\ 
 & 0.400 & $2.16 \pm 0.35^{+0.12}_{-0.14}$ \\ 
 & 0.650 & $0.32 \pm 0.73^{+0.28}_{-0.24}$ \\ 
\hline
\end{tabular}
\end{center}
\caption[]
{The interference structure-function $xF^{\gamma Z}_{3}$
evaluated at $Q^{2}=1\ 500 \gev^{2}$ for $x$ bins centred on $x_c$.
The first (second) error on the measurement
refers to the statistical (systematic) uncertainties.}
\label{tab:xF3gz}
\end{table}
\clearpage
\newpage
\begin{table} \begin{scriptsize} \begin{center} \begin{tabular}[t]{|l|r|l|l|l|r|r|} \hline
\multicolumn{1}{|c|}{$Q^2$ range} & \multicolumn{1}{c|}{$Q^2_c$} & \multicolumn{1}{|c|}{$x$ range} & \multicolumn{1}{c|}{$x_c$} & \multicolumn{1}{c|}{$\tilde{\sigma}$} & \multicolumn{1}{c|}{$N_{\text{data}}$} & \multicolumn{1}{c|}{$N^{\text{MC}}_{\text{bg}}$} \\
\multicolumn{1}{|c|}{($\gev^{2}$)} & \multicolumn{1}{c|}{($\gev^{2}$)} & \multicolumn{1}{|c|}{} & \multicolumn{1}{c|}{} & \multicolumn{1}{c|}{} & & \\ \hline \hline
$185 - 240$ & 200 & $0.004 - 0.006$ & 0.005 & ($1.126 \pm 0.013 ^{+0.029}_{-0.018}$) & 7884 & 61.0 \\ 
 & 200 & $0.006 - 0.010$ & 0.008 & ($0.944 \pm 0.010 ^{+0.008}_{-0.020}$) & 9129 & 22.3 \\ 
 & 200 & $0.010 - 0.017$ & 0.013 & ($0.797 \pm 0.008 ^{+0.003}_{-0.022}$) & 9345 & 8.1 \\ 
 & 200 & $0.017 - 0.025$ & 0.021 & ($0.682 \pm 0.008 ^{+0.013}_{-0.005}$) & 6504 & 3.0 \\ 
 & 200 & $0.025 - 0.037$ & 0.032 & ($0.563 \pm 0.007 ^{+0.015}_{-0.001}$) & 5678 & 0.0 \\ 
 & 200 & $0.037 - 0.060$ & 0.050 & ($0.512 \pm 0.007 ^{+0.011}_{-0.003}$) & 6074 & 0.7 \\
\hline
\end{tabular}
\end{center}
\end{scriptsize}
\caption[]
{The reduced cross-section $\tilde{\sigma}$ for the reaction
$e^{+}p \rightarrow e^{+}X$ ($\mathcal{L} = 78.8 \pbi, P_{e} = +0.32$).
The bin range, bin centre ($Q^2_c$ and $x_c$)
and measured cross section corrected to the electroweak Born level are shown.
Other details as in Table~\ref{tab:dsdq2Total}.
This table has two continuations.}
\label{tab:ds2dxdq2Rh_1}
\end{table}
\begin{table} \begin{scriptsize} \begin{center} \begin{tabular}[t]{|l|r|l|l|l|r|r|} \hline
\multicolumn{1}{|c|}{$Q^2$ range} & \multicolumn{1}{c|}{$Q^2_c$} & \multicolumn{1}{|c|}{$x$ range} & \multicolumn{1}{c|}{$x_c$} & \multicolumn{1}{c|}{$\tilde{\sigma}$} & \multicolumn{1}{c|}{$N_{\text{data}}$} & \multicolumn{1}{c|}{$N^{\text{MC}}_{\text{bg}}$} \\
\multicolumn{1}{|c|}{($\gev^{2}$)} & \multicolumn{1}{c|}{($\gev^{2}$)} & \multicolumn{1}{|c|}{} & \multicolumn{1}{c|}{} & \multicolumn{1}{c|}{} & & \\ \hline \hline
$530 - 710$ & 650 & $0.010 - 0.017$ & 0.013 & ($0.899 \pm 0.018 ^{+0.019}_{-0.025}$) & 2458 & 17.2 \\ 
 & 650 & $0.017 - 0.025$ & 0.021 & ($0.763 \pm 0.019 ^{+0.007}_{-0.018}$) & 1551 & 0.0 \\ 
 & 650 & $0.025 - 0.037$ & 0.032 & ($0.627 \pm 0.018 ^{+0.009}_{-0.025}$) & 1222 & 0.7 \\ 
 & 650 & $0.037 - 0.060$ & 0.050 & ($0.520 \pm 0.015 ^{+0.003}_{-0.010}$) & 1199 & 0.0 \\ 
 & 650 & $0.060 - 0.100$ & 0.080 & ($0.440 \pm 0.013 ^{+0.004}_{-0.011}$) & 1071 & 0.0 \\ 
 & 650 & $0.100 - 0.170$ & 0.130 & ($0.320 \pm 0.011 ^{+0.006}_{-0.003}$) & 892 & 0.0 \\
\hline
\end{tabular}
\end{center}
\end{scriptsize}
\vspace{0.5cm} \hspace{5.4cm} {\bf Table 7:}~~~{\it Continuation 1.}
\label{tab:ds2dxdq2Rh_2}
\end{table}
\begin{table} \begin{scriptsize} \begin{center} \begin{tabular}[t]{|l|r|l|l|l|r|r|} \hline
\multicolumn{1}{|c|}{$Q^2$ range} & \multicolumn{1}{c|}{$Q^2_c$} & \multicolumn{1}{|c|}{$x$ range} & \multicolumn{1}{c|}{$x_c$} & \multicolumn{1}{c|}{$\tilde{\sigma}$} & \multicolumn{1}{c|}{$N_{\text{data}}$} & \multicolumn{1}{c|}{$N^{\text{MC}}_{\text{bg}}$} \\
\multicolumn{1}{|c|}{($\gev^{2}$)} & \multicolumn{1}{c|}{($\gev^{2}$)} & \multicolumn{1}{|c|}{} & \multicolumn{1}{c|}{} & \multicolumn{1}{c|}{} & & \\ \hline \hline
$2500 - 3500$ & 3000  &$0.037 - 0.060$ & 0.050 & ($0.497 \pm 0.029 ^{+0.013}_{-0.012}$) & 290 & 1.3 \\ 
 & 3000 & $0.060 - 0.100$ & 0.080 & ($0.434 \pm 0.024 ^{+0.009}_{-0.009}$) & 338 & 0.0 \\ 
 & 3000 & $0.100 - 0.150$ & 0.130 & ($0.380 \pm 0.023 ^{+0.003}_{-0.015}$) & 271 & 0.0 \\ 
 & 3000 & $0.150 - 0.230$ & 0.180 & ($0.300 \pm 0.020 ^{+0.005}_{-0.004}$) & 225 & 0.0 \\ 
 & 3000 & $0.230 - 0.350$ & 0.250 & ($0.264 \pm 0.020 ^{+0.002}_{-0.008}$) & 181 & 0.0 \\ 
 & 3000 & $0.350 - 0.530$ & 0.400 & ($0.114 \pm 0.014 ^{+0.003}_{-0.003}$) & 70 & 0.0 \\
\hline
\end{tabular}
\end{center}
\end{scriptsize}
\vspace{0.5cm} \hspace{5.4cm} {\bf Table 7:}~~~{\it Continuation 2.}
\label{tab:ds2dxdq2Rh_3}
\end{table}
\clearpage
\newpage
\begin{table} \begin{scriptsize} \begin{center} \begin{tabular}[t]{|l|r|l|l|l|r|r|} \hline
\multicolumn{1}{|c|}{$Q^2$ range} & \multicolumn{1}{c|}{$Q^2_c$} & \multicolumn{1}{|c|}{$x$ range} & \multicolumn{1}{c|}{$x_c$} & \multicolumn{1}{c|}{$\tilde{\sigma}$} & \multicolumn{1}{c|}{$N_{\text{data}}$} & \multicolumn{1}{c|}{$N^{\text{MC}}_{\text{bg}}$} \\
\multicolumn{1}{|c|}{($\gev^{2}$)} & \multicolumn{1}{c|}{($\gev^{2}$)} & \multicolumn{1}{|c|}{} & \multicolumn{1}{c|}{} & \multicolumn{1}{c|}{} & & \\ \hline \hline
$185 - 240$ & 200 & $0.004 - 0.006$ & 0.005 & ($1.089 \pm 0.015 ^{+0.029}_{-0.015}$) & 5429 & 47.1 \\ 
 & 200 & $0.006 - 0.010$ & 0.008 & ($0.947 \pm 0.012 ^{+0.010}_{-0.026}$) & 6518 & 16.5 \\ 
 & 200 & $0.010 - 0.017$ & 0.013 & ($0.807 \pm 0.010 ^{+0.003}_{-0.022}$) & 6729 & 6.0 \\ 
 & 200 & $0.017 - 0.025$ & 0.021 & ($0.679 \pm 0.010 ^{+0.013}_{-0.005}$) & 4603 & 2.1 \\ 
 & 200 & $0.025 - 0.037$ & 0.032 & ($0.571 \pm 0.009 ^{+0.016}_{-0.003}$) & 4089 & 0.0 \\ 
 & 200 & $0.037 - 0.060$ & 0.050 & ($0.510 \pm 0.008 ^{+0.011}_{-0.002}$) & 4301 & 0.5 \\
\hline
\end{tabular}
\end{center}
\end{scriptsize}
\caption[]
{The reduced cross-section $\tilde{\sigma}$ for the reaction
$e^{+}p \rightarrow e^{+}X$ ($\mathcal{L} = 56.7 \pbi, P_{e} = -0.36$).
The bin range, bin centre ($Q^2_c$ and $x_c$)
and measured cross section corrected to the electroweak Born level are shown.
Other details as in Table~\ref{tab:dsdq2Total}.
This table has two continuations.}
\label{tab:ds2dxdq2Lh_1}
\end{table}
\begin{table} \begin{scriptsize} \begin{center} \begin{tabular}[t]{|l|r|l|l|l|r|r|} \hline
\multicolumn{1}{|c|}{$Q^2$ range} & \multicolumn{1}{c|}{$Q^2_c$} & \multicolumn{1}{|c|}{$x$ range} & \multicolumn{1}{c|}{$x_c$} & \multicolumn{1}{c|}{$\tilde{\sigma}$} & \multicolumn{1}{c|}{$N_{\text{data}}$} & \multicolumn{1}{c|}{$N^{\text{MC}}_{\text{bg}}$} \\
\multicolumn{1}{|c|}{($\gev^{2}$)} & \multicolumn{1}{c|}{($\gev^{2}$)} & \multicolumn{1}{|c|}{} & \multicolumn{1}{c|}{} & \multicolumn{1}{c|}{} & & \\ \hline \hline
$530 - 710$ & 650 & $0.010 - 0.017$ & 0.013 & ($0.822 \pm 0.021 ^{+0.033}_{-0.026}$) & 1587 & 11.8 \\ 
 & 650 & $0.017 - 0.025$ & 0.021 & ($0.699 \pm 0.022 ^{+0.010}_{-0.019}$) & 1013 & 0.0 \\ 
 & 650 & $0.025 - 0.037$ & 0.032 & ($0.587 \pm 0.020 ^{+0.010}_{-0.024}$) & 821 & 0.5 \\ 
 & 650 & $0.037 - 0.060$ & 0.050 & ($0.503 \pm 0.017 ^{+0.004}_{-0.011}$) & 829 & 0.0 \\ 
 & 650 & $0.060 - 0.100$ & 0.080 & ($0.426 \pm 0.016 ^{+0.002}_{-0.012}$) & 738 & 0.0 \\ 
 & 650 & $0.100 - 0.170$ & 0.130 & ($0.358 \pm 0.013 ^{+0.003}_{-0.003}$) & 706 & 0.0 \\
\hline
\end{tabular}
\end{center}
\end{scriptsize}
\vspace{0.5cm} \hspace{5.4cm} {\bf Table 8:}~~~{\it Continuation 1.}
\label{tab:ds2dxdq2Lh_2}
\end{table}
\begin{table} \begin{scriptsize} \begin{center} \begin{tabular}[t]{|l|r|l|l|l|r|r|} \hline
\multicolumn{1}{|c|}{$Q^2$ range} & \multicolumn{1}{c|}{$Q^2_c$} & \multicolumn{1}{|c|}{$x$ range} & \multicolumn{1}{c|}{$x_c$} & \multicolumn{1}{c|}{$\tilde{\sigma}$} & \multicolumn{1}{c|}{$N_{\text{data}}$} & \multicolumn{1}{c|}{$N^{\text{MC}}_{\text{bg}}$} \\
\multicolumn{1}{|c|}{($\gev^{2}$)} & \multicolumn{1}{c|}{($\gev^{2}$)} & \multicolumn{1}{|c|}{} & \multicolumn{1}{c|}{} & \multicolumn{1}{c|}{} & & \\ \hline \hline
$2500 - 3500$ & 3000 & $0.037 - 0.060$ & 0.050 & ($0.511 \pm 0.035 ^{+0.012}_{-0.013}$) & 212 & 0.9 \\ 
 & 3000 & $0.060 - 0.100$ & 0.080 & ($0.429 \pm 0.028 ^{+0.006}_{-0.015}$) & 237 & 0.0 \\ 
 & 3000 & $0.100 - 0.150$ & 0.130 & ($0.350 \pm 0.026 ^{+0.005}_{-0.015}$) & 177 & 0.0 \\ 
 & 3000 & $0.150 - 0.230$ & 0.180 & ($0.247 \pm 0.022 ^{+0.005}_{-0.008}$) & 131 & 0.0 \\ 
 & 3000 & $0.230 - 0.350$ & 0.250 & ($0.217 \pm 0.021 ^{+0.003}_{-0.012}$) & 105 & 0.0 \\ 
 & 3000 & $0.350 - 0.530$ & 0.400 & ($0.133 \pm 0.018 ^{+0.006}_{-0.004}$) & 57 & 0.0 \\
\hline
\end{tabular}
\end{center}
\end{scriptsize}
\vspace{0.5cm} \hspace{5.4cm} {\bf Table 8:}~~~{\it Continuation 2.}
\label{tab:ds2dxdq2Lh_3}
\end{table}
\clearpage
\newpage
\begin{table}
\begin{scriptsize}
\begin{center} \begin{tabular}[t]{|c|r|l|r|r|} \hline
\multicolumn{1}{|c|}{$Q^2$ range} & \multicolumn{1}{c|}{$Q^2_c$} & \multicolumn{1}{c|}{$d\sigma / dQ^{2}$} & \multicolumn{1}{c|}{$N_{\text{data}}$} & \multicolumn{1}{c|}{$N^{\text{MC}}_{\text{bg}}$}\\
\multicolumn{1}{|c|}{($\gev^{2}$)} & \multicolumn{1}{c|}{($\gev^{2}$)} & \multicolumn{1}{c|}{($\pb / \gev^{2}$)} & \multicolumn{1}{c|}{} & \multicolumn{1}{c|}{}\\ \hline \hline
$185.0 - 300.0$ & 250 & ($1.07 \pm 0.00 ^{+0.01}_{-0.01}$) $\times$ $10^{1}$ & 96158 & 149.8 \\ 
$300.0 - 400.0$ & 350 & ($4.74 \pm 0.03 ^{+0.07}_{-0.03}$)   & 28667 & 33.6 \\ 
$400.0 - 475.7$ & 440 & ($2.73 \pm 0.03 ^{+0.05}_{-0.05}$)  & 11858 & 36.2 \\ 
$475.7 - 565.7$ & 520 & ($1.83 \pm 0.02 ^{+0.03}_{-0.05}$)  & 8698 & 22.7 \\ 
$565.7 - 672.7$ & 620 & ($1.19 \pm 0.02 ^{+0.02}_{-0.03}$)  & 6108 & 22.4 \\ 
$672.7 - 800.0$ & 730 & ($8.02 \pm 0.11 ^{+0.10}_{-0.24}$) $\times$ $10^{-1}$ & 5859 & 20.0 \\ 
$800.0 - 1050.0$ & 900 & ($4.75 \pm 0.05 ^{+0.07}_{-0.17}$) $\times$ $10^{-1}$ & 7622 & 29.7 \\ 
$1050.0 - 1460.0$ & 1230 & ($2.11 \pm 0.03 ^{+0.05}_{-0.05}$) $\times$ $10^{-1}$ & 5875 & 32.9 \\ 
$1460.0 - 2080.0$ & 1730 & ($9.00 \pm 0.15 ^{+0.23}_{-0.13}$) $\times$ $10^{-2}$ & 3832 & 14.9 \\ 
$2080.0 - 3120.0$ & 2500 & ($3.28 \pm 0.07 ^{+0.07}_{-0.05}$) $\times$ $10^{-2}$ & 2298 & 6.1 \\ 
$3120.0 - 5220.0$ & 3900 & ($1.04 \pm 0.03 ^{+0.02}_{-0.02}$) $\times$ $10^{-2}$ & 1416 & 2.9 \\ 
$5220.0 - 12500.0$ & 7000 & ($1.73 \pm 0.07 ^{+0.02}_{-0.05}$) $\times$ $10^{-3}$ & 616 & 0.0 \\ 
$12500.0 - 51200.0$ & 22400 & ($2.03 \pm 0.26 ^{+0.02}_{-0.08}$) $\times$ $10^{-5}$ & 60 & 0.0 \\ 
\hline
\end{tabular}
\end{center}
\caption[]
{The single-differential cross-section $d\sigma / dQ^{2}$ ($y < 0.9$, $y(1-x)^2>0.004$)
for the reaction $e^{+}p \rightarrow e^{+}X$ ($\mathcal{L} = 78.8 \pbi, P_{e} = +0.32$).
The bin range, bin centre ($Q^2_c$) and measured cross section
corrected to the electroweak Born level are shown.
Other details as in Table~\ref{tab:dsdq2Total}.}
\label{tab:dsdq2Rh}
\end{scriptsize}
\end{table}
\newpage
\begin{table}
\begin{scriptsize}
\begin{center} \begin{tabular}[t]{|r|l|l|l|r|r|} \hline
\multicolumn{1}{|c|}{$Q^2~>$} & \multicolumn{1}{c|}{$x$ range} & \multicolumn{1}{c|}{$x_c$} & \multicolumn{1}{c|}{$d\sigma / dx$} & \multicolumn{1}{c|}{$N_{\text{data}}$} & \multicolumn{1}{c|}{$N^{\text{MC}}_{\text{bg}}$}\\
\multicolumn{1}{|c|}{($\gev^2$)} & \multicolumn{1}{c|}{} & \multicolumn{1}{c|}{} & \multicolumn{1}{c|}{($\pb$)} & & \\
\hline \hline
185 & $(0.63 - 1.00)\times 10^{-2}$ & $0.794\times 10^{-2}$ & ($8.80 \pm 0.06 ^{+0.12}_{-0.13}$) $\times$ $10^{4}$ & 20397 & 91.9 \\ 
 & $(0.10 - 0.16)\times 10^{-1}$ & $0.126\times 10^{-1}$ & ($5.88 \pm 0.04 ^{+0.06}_{-0.17}$) $\times$ $10^{4}$ & 23452 & 71.7 \\ 
 & $(0.16 - 0.25)\times 10^{-1}$ & $0.200\times 10^{-1}$ & ($3.69 \pm 0.02 ^{+0.03}_{-0.03}$) $\times$ $10^{4}$ & 23332 & 49.5 \\ 
 & $(0.25 - 0.40)\times 10^{-1}$ & $0.316\times 10^{-1}$ & ($2.12 \pm 0.01 ^{+0.02}_{-0.01}$) $\times$ $10^{4}$ & 22669 & 17.5 \\ 
 & $(0.40 - 0.63)\times 10^{-1}$ & $0.501\times 10^{-1}$ & ($1.25 \pm 0.01 ^{+0.01}_{-0.01}$) $\times$ $10^{4}$ & 19901 & 3.3 \\ 
 & $(0.63 - 1.00)\times 10^{-1}$ & $0.794\times 10^{-1}$ & ($6.98 \pm 0.05 ^{+0.08}_{-0.03}$) $\times$ $10^{3}$ & 18738 & 2.9 \\
\hline
\end{tabular}
\end{center}
\caption[]
{The single-differential cross-section $d\sigma / dx$ ($y<0.9$, $y(1-x)^2>0.004$)
for $Q^2 > 185 \gev^2$ and $Q^2 > 3\,000 \gev^2$
for the reaction $e^{+}p \rightarrow e^{+}X$ ($\mathcal{L} = 78.8 \pbi,$ $P_{e} = +0.32$).
The $Q^2$ and bin range, bin centre ($x_c$) and measured cross section
corrected to the electroweak Born level are shown.
Other details as in Table~\ref{tab:dsdq2Total}.}
\label{tab:dsdxRh}
\end{scriptsize}
\end{table}
\newpage
\begin{table}
\begin{scriptsize}
\begin{center} \begin{tabular}[t]{|r|r|r|r|r|r|} \hline
\multicolumn{1}{|c|}{$Q^{2}~>$} & \multicolumn{1}{c|}{$y$ range} & \multicolumn{1}{c|}{$y_c$} & \multicolumn{1}{c|}{$d\sigma / dy$} & \multicolumn{1}{c|}{$N_{\text{data}}$} & \multicolumn{1}{c|}{$N^{\text{MC}}_{\text{bg}}$}\\
\multicolumn{1}{|c|}{($\gev^{2}$)} & \multicolumn{1}{c|}{} & \multicolumn{1}{c|}{} & \multicolumn{1}{c|}{($\pb$)} & & \\
\hline \hline
185 & $0.00 - 0.05$ & 0.025 & ($1.65 \pm 0.01 ^{+0.02}_{-0.01}$) $\times$ $10^{4}$ & 44541 & 0.0 \\ 
 & $0.05 - 0.10$ & 0.075 & ($8.19 \pm 0.05 ^{+0.10}_{-0.05}$) $\times$ $10^{3}$ & 29853 & 4.4 \\ 
 & $0.10 - 0.15$ & 0.125 & ($5.73 \pm 0.04 ^{+0.03}_{-0.05}$) $\times$ $10^{3}$ & 20723 & 5.3 \\ 
 & $0.15 - 0.20$ & 0.175 & ($4.40 \pm 0.04 ^{+0.02}_{-0.14}$) $\times$ $10^{3}$ & 15423 & 15.0 \\ 
 & $0.20 - 0.25$ & 0.225 & ($3.66 \pm 0.03 ^{+0.03}_{-0.21}$) $\times$ $10^{3}$ & 12419 & 11.7 \\ 
 & $0.25 - 0.30$ & 0.275 & ($2.96 \pm 0.03 ^{+0.03}_{-0.03}$) $\times$ $10^{3}$ & 9588 & 10.2 \\
\hline
\end{tabular}
\end{center}
\caption[]
{The single-differential cross-section $d\sigma / dy$
for $Q^2 > 185 \gev^2$ and $Q^2 > 3\,000 \gev^2$ ($y(1-x)^2>0.004$)
for the reaction $e^{+}p \rightarrow e^{+}X$ ($\mathcal{L} = 78.8\pbi$, $P_{e} = +0.32$).
The $Q^2$ and bin range, bin centre ($y_c$) and measured cross section
corrected to the electroweak Born level are shown.
Other details as in Table~\ref{tab:dsdq2Total}.}
\label{tab:dsdyRh}
\end{scriptsize}
\end{table}
\newpage
\begin{table}
\begin{scriptsize}
\begin{center} \begin{tabular}[t]{|c|r|l|r|r|} \hline
\multicolumn{1}{|c|}{$Q^2$ range} & \multicolumn{1}{c|}{$Q^2_c$} & \multicolumn{1}{c|}{$d\sigma / dQ^{2}$} & \multicolumn{1}{c|}{$N_{\text{data}}$} & \multicolumn{1}{c|}{$N^{\text{MC}}_{\text{bg}}$}\\
\multicolumn{1}{|c|}{($\gev^{2}$)} & \multicolumn{1}{c|}{($\gev^{2}$)} & \multicolumn{1}{c|}{($\pb / \gev^{2}$)} & \multicolumn{1}{c|}{} & \multicolumn{1}{c|}{}\\ \hline \hline
  $185.0 - 300.0$ & 250 & ($1.06 \pm 0.00 ^{+0.01}_{-0.01}$) $\times$ $10^{1}$ & 67629 & 112.2 \\ 
  $300.0 - 400.0$ & 350 & ($4.55 \pm 0.03 ^{+0.07}_{-0.03}$) & 19579 & 25.5 \\ 
  $400.0 - 475.7$ & 440 & ($2.62 \pm 0.03 ^{+0.05}_{-0.04}$) & 8079 & 27.6 \\ 
  $475.7 - 565.7$ & 520 & ($1.77 \pm 0.02 ^{+0.02}_{-0.05}$) & 5919 & 15.2 \\ 
  $565.7 - 672.7$ & 620 & ($1.11 \pm 0.02 ^{+0.03}_{-0.03}$) & 4058 & 17.0 \\ 
  $672.7 - 800.0$ & 730 & ($7.64 \pm 0.12 ^{+0.19}_{-0.25}$) $\times$ $10^{-1}$ & 3969 & 15.0 \\ 
  $800.0 - 1050.0$ & 900 & ($4.41 \pm 0.06 ^{+0.09}_{-0.16}$) $\times$ $10^{-1}$ & 4994 & 19.7 \\ 
  $1050.0 - 1460.0$ & 1230 & ($2.01 \pm 0.03 ^{+0.04}_{-0.05}$) $\times$ $10^{-1}$ & 3948 & 22.3 \\ 
  $1460.0 - 2080.0$ & 1730 & ($8.25 \pm 0.17 ^{+0.27}_{-0.14}$) $\times$ $10^{-2}$ & 2485 & 10.9 \\ 
  $2080.0 - 3120.0$ & 2500 & ($3.11 \pm 0.08 ^{+0.05}_{-0.06}$) $\times$ $10^{-2}$ & 1541 & 4.4 \\ 
  $3120.0 - 5220.0$ & 3900 & ($8.54 \pm 0.30 ^{+0.15}_{-0.20}$) $\times$ $10^{-3}$ & 826 & 2.1 \\ 
  $5220.0 - 12500.0$ & 7000 & ($1.43 \pm 0.08 ^{+0.02}_{-0.03}$) $\times$ $10^{-3}$ & 361 & 0.0 \\ 
  $12500.0 - 51200.0$ & 22400 & ($1.93 ^{+0.35}_{-0.30}$ $^{+0.11}_{-0.15}$) $\times$ $10^{-5}$ & 41 & 0.0 \\ 
\hline
\end{tabular}
\end{center}
\caption[]
{The single-differential cross-section $d\sigma / dQ^{2}$ ($y < 0.9$, $y(1-x)^2>0.004$)
for the reaction $e^{+}p \rightarrow e^{+}X$ ($\mathcal{L} = 56.7 \pbi, P_{e} = -0.36$).
The bin range, bin centre ($Q^2_c$) and measured cross section
corrected to the electroweak Born level are shown.
Other details as in Table~\ref{tab:dsdq2Total}.}
\label{tab:dsdq2Lh}
\end{scriptsize}
\end{table}
\newpage
\begin{table}
\begin{scriptsize}
\begin{center} \begin{tabular}[t]{|r|l|l|l|r|r|} \hline
\multicolumn{1}{|c|}{$Q^2~>$} & \multicolumn{1}{c|}{$x$ range} & \multicolumn{1}{c|}{$x_c$} & \multicolumn{1}{c|}{$d\sigma / dx$} & \multicolumn{1}{c|}{$N_{\text{data}}$} & \multicolumn{1}{c|}{$N^{\text{MC}}_{\text{bg}}$}\\
\multicolumn{1}{|c|}{($\gev^2$)} & \multicolumn{1}{c|}{} & \multicolumn{1}{c|}{} & \multicolumn{1}{c|}{($\pb$)} & & \\
\hline \hline
185 & $(0.63 - 1.00)\times 10^{-2}$ & $0.794\times 10^{-2}$ & ($8.61 \pm 0.07 ^{+0.16}_{-0.16}$) $\times$ $10^{4}$ & 14173 & 69.1 \\ 
 & $(0.10 - 0.16)\times 10^{-1}$ & $0.126\times 10^{-1}$ & ($5.79 \pm 0.05 ^{+0.07}_{-0.17}$) $\times$ $10^{4}$ & 16410 & 50.8 \\ 
 & $(0.16 - 0.25)\times 10^{-1}$ & $0.200\times 10^{-1}$ & ($3.54 \pm 0.03 ^{+0.03}_{-0.03}$) $\times$ $10^{4}$ & 15901 & 33.4 \\ 
 & $(0.25 - 0.40)\times 10^{-1}$  & $0.316\times 10^{-1}$ & ($2.07 \pm 0.02 ^{+0.02}_{-0.01}$) $\times$ $10^{4}$ & 15715 & 12.7 \\ 
 & $(0.40 - 0.63)\times 10^{-1}$ & $0.501\times 10^{-1}$ & ($1.21 \pm 0.01 ^{+0.01}_{-0.01}$) $\times$ $10^{4}$ & 13656 & 2.3 \\ 
 & $(0.63 - 1.00)\times 10^{-1}$ & $0.794\times 10^{-1}$ & ($6.84 \pm 0.06 ^{+0.08}_{-0.03}$) $\times$ $10^{3}$ & 13087 & 2.1 \\
\hline
\end{tabular}
\end{center}
\caption[]
{The single-differential cross-section $d\sigma / dx$ ($y<0.9$, $y(1-x)^2>0.004$)
for $Q^2 > 185 \gev^2$ and $Q^2 > 3\,000 \gev^2$
for the reaction $e^{+}p \rightarrow e^{+}X$ ($\mathcal{L} = 56.7 \pbi,$ $P_{e} = -0.36$).
The $Q^2$ and bin range, bin centre ($x_c$) and measured cross section
corrected to the electroweak Born level are shown.
Other details as in Table~\ref{tab:dsdq2Total}.}
\label{tab:dsdxLh}
\end{scriptsize}
\end{table}
\newpage
\begin{table}
\begin{scriptsize}
\begin{center} \begin{tabular}[t]{|r|r|r|r|r|r|} \hline
\multicolumn{1}{|c|}{$Q^{2}~>$} & \multicolumn{1}{c|}{$y$ range} & \multicolumn{1}{c|}{$y_c$} & \multicolumn{1}{c|}{$d\sigma / dy$} & \multicolumn{1}{c|}{$N_{\text{data}}$} & \multicolumn{1}{c|}{$N^{\text{MC}}_{\text{bg}}$}\\
\multicolumn{1}{|c|}{($\gev^{2}$)} & \multicolumn{1}{c|}{} & \multicolumn{1}{c|}{} & \multicolumn{1}{c|}{($\pb$)} & & \\
\hline \hline
185 & $0.00 - 0.05$ & 0.025 & ($1.60 \pm 0.01 ^{+0.02}_{-0.01}$) $\times$ $10^{4}$ & 30773 & 0.0 \\ 
 & $0.05 - 0.10$ & 0.075 & ($7.98 \pm 0.06 ^{+0.09}_{-0.04}$) $\times$ $10^{3}$ & 20679 & 3.0 \\ 
 & $0.10 - 0.15$ & 0.125 & ($5.53 \pm 0.05 ^{+0.03}_{-0.04}$) $\times$ $10^{3}$ & 14221 & 6.3 \\ 
 & $0.15 - 0.20$ & 0.175 & ($4.34 \pm 0.04 ^{+0.03}_{-0.14}$) $\times$ $10^{3}$ & 10814 & 9.7 \\ 
 & $0.20 - 0.25$ & 0.225 & ($3.55 \pm 0.04 ^{+0.03}_{-0.20}$) $\times$ $10^{3}$ & 8555 & 8.1 \\ 
 & $0.25 - 0.30$ & 0.275 & ($2.89 \pm 0.04 ^{+0.05}_{-0.05}$) $\times$ $10^{3}$ & 6666 & 7.6 \\
\hline
\end{tabular}
\end{center}
\caption[]
{The single-differential cross-section $d\sigma / dy$
for $Q^2 > 185 \gev^2$ and $Q^2 > 3\,000 \gev^2$ ($y(1-x)^2>0.004$)
for the reaction $e^{+}p \rightarrow e^{+}X$ ($\mathcal{L} = 56.7\pbi$, $P_{e} = -0.36$).
The $Q^2$ and bin range, bin centre ($y_c$) and measured cross section
corrected to the electroweak Born level are shown.
Other details as in Table~\ref{tab:dsdq2Total}.}
\label{tab:dsdyLh}
\end{scriptsize}
\end{table}
\newpage
\begin{table}
\begin{center}
\begin{tabular}[t]{|c|r|l|l|} \hline
\multicolumn{1}{|c|}{$Q^{2}$ range} & \multicolumn{1}{c|}{$Q^{2}_c$} & \multicolumn{1}{c|}{Ratio}& \multicolumn{1}{c|}{Asymmetry $A^{+}$} \\
\multicolumn{1}{|c|}{($\gev^{2}$)} & \multicolumn{1}{c|}{($\gev^{2}$)} &  & \multicolumn{1}{c|}{$\times 10$} \\ \hline \hline
$185.0 - 300.0 $  &  250  &  $1.01\pm 0.01$ &  $0.18\pm 0.07$  \\ 
$300.0 - 400.0 $  &  350  &  $1.04\pm 0.01$ &  $0.57\pm 0.14$  \\ 
$400.0 - 475.7 $  &  440  &  $1.04\pm 0.02$ &  $0.59\pm 0.21$  \\ 
$475.7 - 565.7 $  &  520  &  $1.04\pm 0.02$ &  $0.54\pm 0.25$  \\ 
$565.7 - 672.7 $  &  620  &  $1.07\pm 0.02$ &  $0.95\pm 0.30$  \\
$672.7 - 800. $  &  730  &  $1.05\pm 0.02$ &  $0.71\pm 0.30$  \\ 
$800.0 - 1050.0 $  &  900  &  $1.08\pm 0.02$ &  $1.09\pm 0.27$  \\ 
$1050.0 - 1460.0 $  &  1230  &  $1.05\pm 0.02$ &  $0.73\pm 0.30$  \\ 
$1460.0 - 2080.0 $  &  1730  &  $1.09\pm 0.03$ &  $1.27\pm 0.38$  \\ 
$2080.0 - 3120.0 $  &  2500  &  $1.05\pm 0.03$ &  $0.77\pm 0.48$  \\ 
$3120.0 - 5220.0 $  &  3900  &  $1.21\pm 0.05$ &  $2.82\pm 0.63$  \\ 
$5220.0 - 12500.0 $  &  7000  &  $1.21\pm 0.08$ &  $2.74\pm 0.95$  \\ 
$12500.0 - 51200.0 $  &  22400  &  $1.05\pm 0.21$ &  $0.76\pm 2.96$  \\
\hline
\end{tabular}
\end{center}
\caption[]
{The polarisation asymmetry measured using
positively and negatively polarised $e^{+}p$ beams
($\mathcal{L} = 78.8 \pbi, P_{e} = +0.32$ and
 $\mathcal{L} = 56.7 \pbi, P_{e} = -0.36$, respectively).
The bin range, bin centre ($Q^2_c$), the cross section ratio of the samples with $P_{e} = +0.32$ and $P_{e} = -0.36$ and the measured asymmetry $A^+$ are shown.
Only the statistical uncertainties on the measurement are shown
as systematic uncertainties are assumed to cancel.}
\label{tab:asym}
\end{table}


%
%
\end{document}